\documentclass[fleqn,usenatbib]{mnras}

\usepackage{newtxtext,newtxmath}
\usepackage[T1]{fontenc}

\DeclareRobustCommand{\VAN}[3]{#2}
\let\VANthebibliography\thebibliography
\def\thebibliography{\DeclareRobustCommand{\VAN}[3]{##3}\VANthebibliography}

\usepackage{graphicx}	% Including figure files
\usepackage{amsmath}	% Advanced maths commands
\usepackage{bm}
\usepackage{array}

\newcommand{\be}{\begin{equation}}
\newcommand{\ee}{\end{equation}}
\newcommand{\bea}{\begin{eqnarray}}
\newcommand{\eea}{\end{eqnarray}}
\newcommand{\dd}{\mathrm{d}}

\title[Feedback from marginally-resolved HII regions]{Momentum feedback from marginally-resolved HII regions in isolated disc galaxies}

\author[S.~M.~R.~Jeffreson et al.]{
S.~M.~R.~Jeffreson,$^{1,2}$\thanks{sarah.jeffreson@cfa.harvard.edu}
Mark~R.~Krumholz$^{3,4}$, Yusuke~Fujimoto$^{5,6}$, Lucia Armillotta$^{7}$,
\newauthor
Benjamin~W.~Keller$^{2}$, M{\'elanie}~Chevance$^{2}$, J.~M.~Diederik~Kruijssen$^{2}$
\\
$^{1}$ Center for Astrophysics, Harvard \& Smithsonian, 60 Garden St, Cambridge, MA 02138, USA \\
$^{2}$ Astronomisches Rechen-Institut, Zentrum f\"{u}r Astronomie der Universit\"{a}t Heidelberg, M\"{o}nchhofstra\ss e 12-14, 69120 Heidelberg, Germany \\
$^{3}$ Research School of Astronomy and Astrophysics, Australian National University, Canberra, ACT 2611 Australia \\
$^{4}$ ARC Centre of Excellence for Astronomy in Three Dimensions (ASTRO-3D), Canberra, ACT 2611 Australia \\
$^{5}$ Earth and Planets Laboratory, Carnegie Institution for Science, 5241 Broad Branch Road, NW, Washington, DC 20015, USA\\
$^{6}$ Department of Astrophysical Sciences, Princeton University, Princeton, NJ 08544, USA \\
}

\date{Accepted 2021 May 24. Received 2021 March 31; in original form 2021 March 31}

\pubyear{2021}

\begin{document}
\label{firstpage}
\pagerange{\pageref{firstpage}--\pageref{lastpage}}
\maketitle

\begin{abstract}
We present a novel, physically-motivated sub-grid model for HII region feedback within the moving mesh code {\sc Arepo}, accounting for both the radiation pressure-driven and thermal expansion of the ionised gas surrounding young stellar clusters. We apply this framework to isolated disc galaxy simulations with mass resolutions between $10^3~{\rm M}_\odot$ and $10^5~{\rm M}_\odot$ per gas cell. Each simulation accounts for the self-gravity of the gas, the momentum and thermal energy from supernovae, the injection of mass by stellar winds, and the non-equilibrium chemistry of hydrogen, carbon and oxygen. We reduce the resolution-dependence of our model by grouping those HII regions with overlapping ionisation front radii. The Str\"{o}mgren radii of the grouped HII regions are at best marginally-resolved, so that the injection of purely-thermal energy within these radii has no effect on the interstellar medium. By contrast, the injection of momentum increases the fraction of cold and molecular gas by more than 50~per~cent at mass resolutions of $10^3~{\rm M}_\odot$, and decreases its turbulent velocity dispersion by $\sim 10~{\rm kms}^{-1}$. The mass-loading of galactic outflows is decreased by an order of magnitude. The characteristic lifetime of the least-massive molecular clouds ($M/{\rm M}_\odot \la 5.6 \times 10^4$) is reduced from $\sim 18$~Myr to $\la 10$~Myr, indicating that HII region feedback is effective in destroying these clouds. Conversely, the lifetimes of intermediate-mass clouds ($5.6 \times 10^4 \la M/{\rm M}_\odot \la 5 \times 10^5$) are elongated by $\sim 7$~Myr, likely due to a reduction in supernova clustering. The derived cloud lifetimes span the range from $10$-$40$~Myr, in agreement with observations. All results are independent of whether the momentum is injected from a ‘spherical’ or a ‘blister-type’ HII region.
\end{abstract}

%%%%%%%%%%%%%%%%%%%%%%%%%%%%%%%%%%%%%%%%%%%%%%%%%%

%%%%%%%%%%%%%%%%% BODY OF PAPER %%%%%%%%%%%%%%%%%%

\section{Introduction}
\label{Sec::Introduction}
Stellar feedback is a crucial feature of realistic galaxy simulations, from the cosmological scales of galaxy formation and evolution down to the sub-galactic scales of molecular cloud formation and dispersal. Comparable quantities of energy and momentum are injected into the interstellar medium by the explosive destruction of stars at the end of their lifetimes (supernovae), and by the expansion of the heated, ionised gas surrounding massive stars~\citep[HII regions, see][for details]{Matzner02,2013MNRAS.432..455D}. Smaller but significant quantities are injected by the outflows of gas generated within the stellar atmosphere (stellar winds). This energy and momentum is responsible for the formation of a three-phase interstellar medium~\citep{1977ApJ...218..148M}, for the generation of galactic outflows~\citep[e.g.][]{2014MNRAS.442.3013K,2020ApJ...903L..34K}, and for the dispersal of molecular gas surrounding young stars~\citep[e.g.][]{2019Natur.569..519K,2020arXiv201013788C,Chevance20}. Its interplay with gravity determines the gas disc scale-heights of galaxies and affects the turbulent velocity dispersions of their cold neutral media and molecular components~\citep{OstrikerShetty2011}. As such, stellar feedback plays an important role in limiting the star formation efficiencies of individual molecular clouds, and so the star formation rates of entire galaxies.

The implementation of supernova feedback in galactic and cosmological simulations has historically presented a number of problems. The inefficient conversion of thermal energy to momentum when the momentum-generating phase of blast-wave expansion is unresolved (the `overcooling problem') leads to an underestimation of the kinetic energy injected into the interstellar medium~\citep[e.g.][]{Katz92,1993MNRAS.265..271N,2003MNRAS.339..289S,2006MNRAS.371.1125S}. Overcooling can be most-effectively resolved by explicitly injecting the terminal momentum of the blast-wave~\citep[e.g.][]{KimmCen14}, and several different parametrisations of the terminal momentum are available, derived from high-resolution simulations of supernova remnants~\citep[e.g.][]{Gentry17,2020MNRAS.492.1243G}. Within clusters of stars, the distribution of supernova frequencies, energies and `delay times' between stellar birth and the first supernova detonation depends sensitively on the massive end of the initial stellar mass function (IMF) from which the stellar population is drawn~\citep{2020arXiv200911309S,2020arXiv201010533S,2020arXiv200403608K}. Variations in the methods used to stochastically-sample the IMF~\citep[e.g.][]{2010A&A...512A..79H,2013NewAR..57..123C,daSilva14,Krumholz15,2020arXiv201010533S} can therefore introduce large variations in the final energy and momentum injected by supernova feedback. For example, erroneously-long delays will lead to overly-violent supernova explosions due to an increased degree of supernova clustering, as well as to the overly-slow removal of gas from around young stars~\citep[e.g.][]{Fujimoto19}. These two factors will directly influence the giant molecular cloud lifetime, the star formation efficiency within individual clouds, and so the star formation efficiencies of galaxies.

In simulations of entire galaxies~\citep{2013MNRAS.428..129S,Agertz13,Hopkins18,2019MNRAS.489.4233M,2020arXiv200911309S}, supernova feedback on its own has been found inadequate to reproduce the observed properties of the three-phase interstellar medium. This has led to the inclusion of `pre-supernova feedback' mechanisms in most such state-of-the-art hydrodynamical simulations, which inject energy and momentum into the interstellar medium before the detonation of the first supernovae. The mechanisms these recipes are intended to model include the energy provided by direct and dust-reprocessed radiation pressure, stellar winds and photoionisation. The explicit injection of momentum from pre-supernova feedback, at mass resolutions between $10^3$ and $10^4$ solar masses, is found to make a qualitative difference to the large-scale properties of the interstellar medium~\citep[e.g.][]{2013MNRAS.434.3142A,Agertz13,2015ApJ...804...18A,2020arXiv200911309S}. However, it is likely that these treatments substantially over-estimate the effectiveness of direct radiation pressure in driving gas away from young stars, because they do not resolve the radius of the dust destruction front~\citep{2018MNRAS.480.3468K}. In isolated disc galaxy simulations including purely-thermal energy injection from HII regions~\citep[e.g.][]{Goldbaum16,Fujimoto19}, pre-supernova feedback is unable to drive dense gas away from young star particles, or to destroy molecular clouds, implying that HII region feedback may suffer from the same `overcooling problem' as does supernova feedback, if the Str\"{o}mgren radii of the HII regions are unresolved or only marginally-resolved.

Disentangling the relative roles of supernovae and pre-supernova feedback mechanisms from the spurious numerical variations of each individual feedback mechanism is a work in progress. Here we examine the relative effects of HII region feedback and supernova feedback when the Str\"{o}mgren radii of the HII regions are marginally-resolved, at mass resolutions spanning the range from $10^3$ down to $10^5~{\rm M}_\odot$ per gas cell. To this end, we develop a novel, physically-motivated sub-grid model for the thermal energy and momentum injected by HII regions at these resolutions, based on the analytic theory of~\cite{KrumholzMatzner09}. This explicitly includes the momentum from both radiation and gas pressure, and we ensure that our prescription is as close to convergence as possible, by introducing a grouping prescription for simulated star particles with overlapping ionisation front radii. We investigate the effect of our feedback prescription on the interstellar medium and on its giant molecular cloud population, relative to the case of supernova feedback only. We also test the effect of thermal vs.~kinetic energy injection from the HII regions, and of the influence of HII-region shape (spherical, or `blister-type', sitting at the edge of a giant molecular cloud).

The remainder of the paper is organised as follows. Our `control' isolated galaxy simulation, without HII region feedback, is described in Section~\ref{Sec::SNe-only}. The analytic theory and numerical implementation of our HII region feedback model is detailed in Section~\ref{Sec::HII-region-fb}. This section also includes convergence tests for the individual components of the model. In Section~\ref{Sec::results} we describe the impact of HII region feedback on the large-scale properties of the interstellar medium and on the cloud-scale properties of its cold molecular gas, across a range of mass resolutions from $10^3$ to $10^5~{\rm M}_\odot$ per gas cell. Section~\ref{Sec::discussion} describes the caveats of our model and compares our results to existing studies in the literature. Finally, a summary of our conclusions is given in Section~\ref{Sec::conclusion}.

\section{Basic simulation setup (no HII region feedback)} \label{Sec::SNe-only}
We run the simulations presented in this work using the moving-mesh hydrodynamics code {\sc Arepo}~\citep{Springel10}. Within {\sc Arepo}, the gaseous component of each simulation is modelled by an unstructured moving mesh, defined by the Voronoi tessellation about a discrete set of points. The mesh moves with the fluid flow in a way similar to that of smoothed particle hydrodynamics codes, however its spatial resolution can be refined locally and automatically to arbitrarily-high levels without requiring large over-densities to be present in the regions of interest, as in Eulerian codes. As such, a high level of accuracy can be maintained in dealing with shock fronts and low-density flows. 

We use the isolated disc initial condition generated for the Agora comparison project~\citep{Kim14}. This initial condition is designed to resemble a Milky Way-like galaxy at redshift $z \sim 0$. It has a~\cite{Navarro97} dark matter halo of mass $M_{200} = 1.07 \times 10^{12}~{\rm M}_\odot$, a virial radius of $R_{200} = 205~{\rm kpc}$, a halo concentration parameter of $c = 10$ and a spin parameter of $\lambda = 0.04$. The stellar bulge is of~\cite{Hernquist90} type and has a mass of $3.437 \times 10^9~{\rm M}_\odot$, while the exponential disc has a mass of $4.297 \times 10^{10}~{\rm M}_\odot$, a scale-length of $3.43~{\rm kpc}$, and a scale-height of $0.34~{\rm kpc}$. The bulge to stellar disc ratio is 0.125 and the overall gas fraction is $0.18$. The lowest-resolution disc (LOW) has a star particle mass (for the initial disc and bulge) of $3.437 \times 10^5~{\rm M}_\odot$, and a dark matter particle mass of $1.254 \times 10^7~{\rm M_\odot}$. The medium-resolution (MED) and high-resolution (HI) discs have stellar and dark matter particle masses that are smaller by factors of $0.1$ and $0.01$, respectively. All discs were generated using the MakeNewDisc code~\citep{Springel05}. During runtime, we set a softening length of $80~{\rm pc}$ for the LOW disc, $40~{\rm pc}$ for the MED disc and $20~{\rm pc}$ for the HI disc, with median gas cell masses of $8.59 \times 10^4~{\rm M}_\odot$, $8.59 \times 10^3~{\rm M}_\odot$ and $859~{\rm M}_\odot$ respectively, following~\cite{Kim14}. We set the initial gas temperature to be $10^4~{\rm K}$, however this re-equilibriates within the first $\leq 10~{\rm Myr}$ of run-time, according to the state of thermal balance between line-emission cooling and heating due to photoelectric emission from dust grains and polycyclic aromatic hydrocarbons (PAHs), as modelled by our chemical network.

Our star formation prescription locally reproduces the observed relation between the star formation rate and gas surface density~\citep{Kennicutt98}. The star formation rate density in a gas cell $i$ with volume density $\rho_i$ is given by
\begin{equation}
\label{Eqn::starformation}
\frac{\dd \rho_{*,i}}{\dd t} = 
\begin{cases}
      \frac{\epsilon_{\rm ff} \rho_i}{t_{{\rm ff},i}}, \; \rho_i \geq \rho_{\rm thres} \\
      0, \; \rho_i < \rho_{\rm thres}\\
   \end{cases},
\end{equation}
where $t_{{\rm ff}, i} = \sqrt{3\pi/(32 G\rho_i)}$ is the local free-fall time-scale and we set the density threshold $\rho_{\rm thresh}$ for star formation at $1~{\rm cm}^{-3}$, $100~{\rm cm}^{-3}$ and $1000~{\rm cm}^{-3}$ for the LOW, MED and HI discs, respectively. Our choice of $\rho_{\rm thresh}$ ensures that the densest gas in each simulation is Jeans-unstable at the corresponding mass resolution. We set the star formation efficiency at $\epsilon_{\rm ff} = 0.1$, consistent with the upper end of the observed range in dense, molecular gas~\citep{1974ApJ...192L.149Z,Krumholz&Tan07,2012ApJ...745...69K,2014ApJ...782..114E,2016A&A...588A..29H}. We also tested a value of $\epsilon_{\rm ff} = 0.01$ at the lower end of the observed range, however found that this produced much higher turbulent velocity dispersions in the galactic mid-plane of up to $40~{\rm km~s}^{-1}$, inconsistent with observations of the Milky Way's disc~\citep[e.g.][]{2009AJ....137.4424T}.

We follow the stellar evolution of each live star particle formed during the simulation using the Stochastically Lighting up Galaxies code (SLUG), described in~\cite{daSilva12,daSilva14,Krumholz15}. Each star particle of birth mass $m_{\rm birth}$ is assigned a `cluster’ that is populated with $N$ stars drawn randomly from a~\cite{2005ASSL..327...41C} initial stellar mass function (IMF), where $N$ is chosen from a Poisson distribution with an expectation value of $m_{\rm birth}/\overline{M}$, with $\overline{M}$ the mean mass of a single star. For a large number of star particles, this procedure ensures that the IMF is fully-sampled across the stellar population, and that the total mass of populated stars equals the combined mass of the star particles. Within each cluster, SLUG evolves individual stars along Padova solar metallicity tracks~\citep{Fagotto94a,Fagotto94b,VazquezLeitherer05} with starburst99-like spectral synthesis~\citep{Leitherer99}.

In addition to the model for HII region feedback described in the following section, we include feedback from supernovae and mass ejection from stellar winds. Our stellar evolution model provides the number of supernovae $N_{*, {\rm SN}}$ produced by every star particle, along with the mass loss $\Delta m_*$ due either to supernovae or to stellar winds. For $N_{*, {\rm SN}} = 0$, we treat the mass loss as resulting from stellar winds, and simply deposit $\Delta m_*$ into the nearest gas cell. If $N_{*, {\rm SN}} > 0$, we make the approximation that all mass loss results from supernovae. We do not resolve the energy-conserving, momentum-generating phase of supernova blast-wave expansion in our simulations, such that we must calculate the terminal momentum of the blast-wave explicitly to prevent over-cooling, following the prescription of~\cite{KimmCen14}. We use the (unclustered) parametrisation of the terminal momentum injected into the gas cells $k$ neighbouring a central cell $j$, derived from the high-resolution simulations of~\cite{Gentry17}, and given by
 \begin{equation} \label{Eqn::gentry17}
 \frac{p_{{\rm t}, k}}{{\rm M}_\odot {\rm kms}^{-1}} = 4.249 \times 10^5 N_{j, {\rm SN}} \Big(\frac{n_k}{{\rm cm}^{-3}}\Big)^{-0.06},
 \end{equation}
where $N_{j, {\rm SN}}$ is the cumulative number of supernovae received by a gas cell $j$ from all of the star particles for which it is the nearest neighbour. This terminal momentum is then spread into the cells surrounding the central cell, as in~\cite{2020arXiv200403608K,2020MNRAS.498..385J,Jeffreson21a}, with an upper limit set by kinetic energy conservation as the shell sweeps up the mass in the cells surrounding the central one~\citep[see also][for similar prescriptions]{Hopkins18,Smith2018}. A convergence test for a single supernova explosion, implemented via the above method, is presented in Appendix~\ref{App::res-tests}.

The chemical composition of the gas in our simulations evolves according to the simplified network of hydrogen, carbon and oxygen chemistry described in~\cite{GloverMacLow07a,GloverMacLow07b} and in~\cite{NelsonLanger97}. For each Voronoi gas cell, fractional abundances are computed and tracked for the chemical species ${\rm H}$, ${\rm H}_2$, ${\rm H}^+$, ${\rm He}$, ${\rm C}^+$, ${\rm CO}$, ${\rm O}$ and ${\rm e}^-$. The chemistry is coupled to the heating and cooling of the interstellar medium via the atomic and molecular cooling function of~\cite{Glover10}. The full list of heating and cooling processes is given in their Table 1. As such, the heating and cooling rates in our simulations depend not only on the gas density and temperature, but also on the strength of the interstellar radiation field, the cosmic-ray ionisation rate, the dust fraction and temperature, and on the set of chemical abundances tracked for each gas cell. We assign a value of $1.7$~Habing fields to the UV component of the ISRF according to~\cite{Mathis83}, a value of $3 \times 10^{-17}$~s$^{-1}$ to the cosmic ionisation rate~\citep{2000A&A...358L..79V}, and assume the solar value for the dust-to-gas ratio.

\section{HII region feedback} \label{Sec::HII-region-fb}
In this section, we derive the momentum per unit time provided by a single HII region to the surrounding interstellar medium. We develop a novel sub-grid model for injecting this momentum in simulations that do not resolve the median Str{\"o}mgren radius. We also describe our prescription for heating the interstellar medium within this radius. Parts of the following prescription are used in the simulations `HII heat', `HII spherical mom.', `HII beamed mom.' and `HII heat \& beamed mom.', as listed in Table~\ref{Tab::sims}.

\begin{table}
\begin{center}
\label{Tab::sims}
  \caption{The five simulations run in this work and their feedback prescriptions.}
  \begin{tabular}{@{}l m{4.5cm}@{}}
  \hline
   Simulation name & Feedback prescription \\
  \hline
   SNe (control) & Supernovae only \\
   HII heat & Supernovae plus thermal HII regions \\
   HII spherical mom. & Supernovae plus spherical HII region momentum \\
   HII beamed mom. & Supernovae plus beamed HII region momentum \\
   HII heat \& beamed mom. & Supernovae plus thermal HII regions plus beamed HII region momentum \\
  \hline
\end{tabular}
\end{center}
\end{table}
\subsection{Analytic theory} \label{Sec::Theory}
We consider the momentum injected into a molecular cloud by a massive star or stellar cluster that produces ionising photons of energy $\epsilon_0 \sim 13.6$~eV at a rate $S$. We choose a system of co-ordinates that is centred on the ionising source, and following~\citealt{KrumholzMatzner09} (hereafter KM09), we parametrize the density profile of the surrounding gas as a power-law of the form $\rho(r) = \rho_0(r/r_0)^{-k_\rho}$. A source may be fully `embedded' within the host cloud, such that it is surrounded on all sides by dense gas, or it may be located at the edge of a molecular cloud, such that it produces a `blister-type', hemispherical HII region, for which we take $\rho = 0$ for the outward-facing hemisphere. The photons from the source transfer their energy to the surrounding gas via two primary mechanisms. Firstly, kinetic energy is carried away by the products of ionisation (free electrons, hydrogen nuclei and helium nuclei), heating the ionised material to a temperature of $T_{\rm II}$~\citep{Spitzer78}. As the sound speed $c_{\rm II}$ in the ionised gas is much higher than that in the surrounding neutral gas, the initial expansion of the HII region sweeps up a thin shell of neutral material that separates the ionised region from its surroundings. Secondly, photons may be absorbed by dust grains and hydrogen atoms, delivering a momentum kick that accelerates the particles away from the ionising source. The radiative acceleration will always be highest closest to the source (KM09), again contributing to the production of a thin shell bounding the HII region. As such, the momentum delivered to the dense gas outside the HII region is very well approximated by the momentum of this bounding shell. The momentum equation for this shell may be written in the form of~\cite{Matzner02} as
\begin{equation}
\label{Eqn::momentumequation_orig}
\frac{\dd p}{\dd t} = \frac{\dd}{\dd t}(M_{\rm sh} \dot{r}_{\rm II}) = A_{\rm sh}\Big[P_{\rm gas} + P_{\rm rad}\Big],
\end{equation}
where $M_{\rm sh} = (2,4)\pi r_{\rm II}^3 \overline{\rho}(r_{\rm II})$ is the mass of neutral material swept into the shell of ionisation front radius $r_{\rm II}$ during its initial rapid expansion, $A_{\rm sh} = (2,4)\pi r_{\rm II}^2$ is its surface area, and $\overline{\rho}(r) = 3/(3-k_\rho) \rho_0 (r/r_0)^{-k_\rho}$ is the mean volume density inside a radius $r$ in the initial molecular cloud. The first value in the above parentheses corresponds to the case of a blister-type HII region, in which the gas pressure at the HII-cloud interface is augmented by a thrust of equal magnitude and direction, due to the flux of gas through the opposing hemisphere.  The equality of the pressure and thrust terms depends on the assumption that the ejected gas can escape freely from the HII region, such that its velocity relative to the velocity $\dot{r}_{\rm II}$ of the ionisation front tends towards the speed of sound within the HII region~\citep{Kahn54}. The second value in parentheses then corresponds to the case of an embedded HII region, in which no thrust is produced. The momenta delivered by thermal heating and radiative acceleration are given in terms of a gas pressure $P_{\rm gas}$ and a radiation pressure $P_{\rm rad}$, respectively. The gas pressure is given by
\begin{equation}
\label{Eqn::Pgas}
P_{\rm gas} = (2,1) \rho_{\rm II} c_{\rm II}^2,
\end{equation}
where $\rho_{\rm II}$ refers to the density of the heated, ionised gas inside the swept-up shell. The radiation pressure term in Equation (\ref{Eqn::momentumequation_orig}) can be written in the form of KM09 as
\begin{equation}
\label{Eqn::Prad}
P_{\rm rad} = \frac{f_{\rm trap} \epsilon_0 S}{4\pi c r_{\rm II}^2},
\end{equation}
where the factor $f_{\rm trap}$ quantifies the enhancement of the radiative force via the trapping of photons and stellar winds in the expanding shell, and $c$ is the speed of light.

% ==================== GROUPING FIGURES
\begin{figure*}
  \label{Fig::FoF-convergence-test}
    \includegraphics[width=\linewidth]{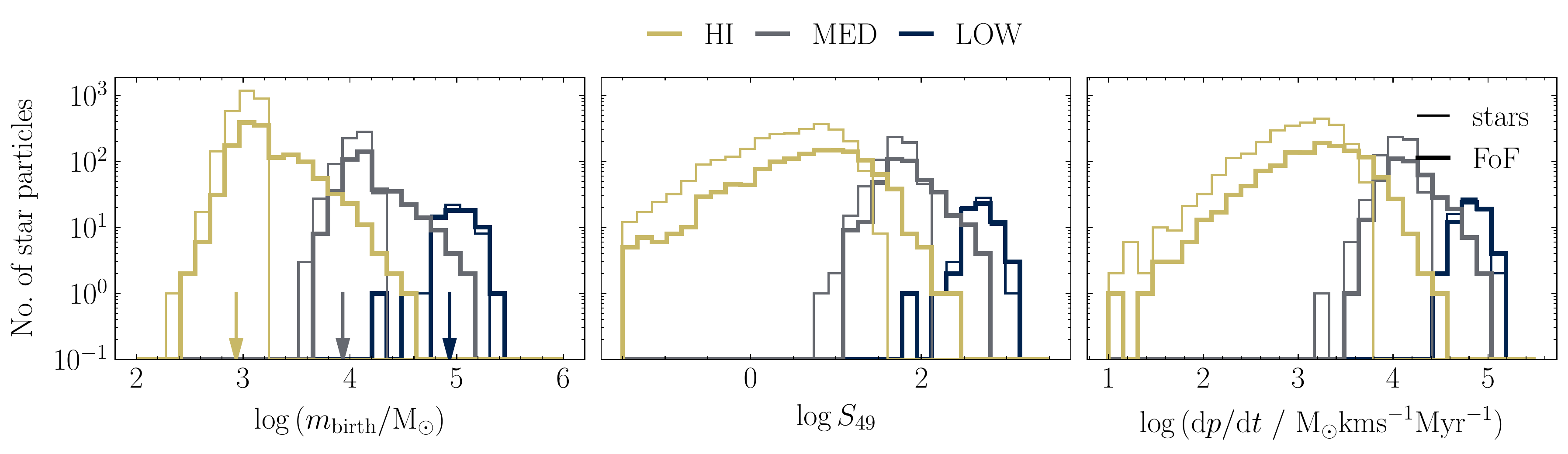}
    \caption{Distributions of birth masses (left), ionising luminosities (centre) and momentum injection rates (right) for the star particles (thin lines) and FoF groups of star particles with overlapping ionisation front radii (bold lines) in our three `HII heat \& beamed mom.' simulations at time $t=600$~Myr. Each line colour corresponds to a different numerical mass resolution: $10^5~{\rm M}_\odot$ (LOW, dark blue), $10^4~{\rm M}_\odot$ (MED, grey), and $10^3~{\rm M}_\odot$ (HI, yellow). The arrows in the left-hand panel point to the median mass resolution in each galaxy.}
\end{figure*}

\begin{figure}
  \label{Fig::rst0}
    \includegraphics[width=\linewidth]{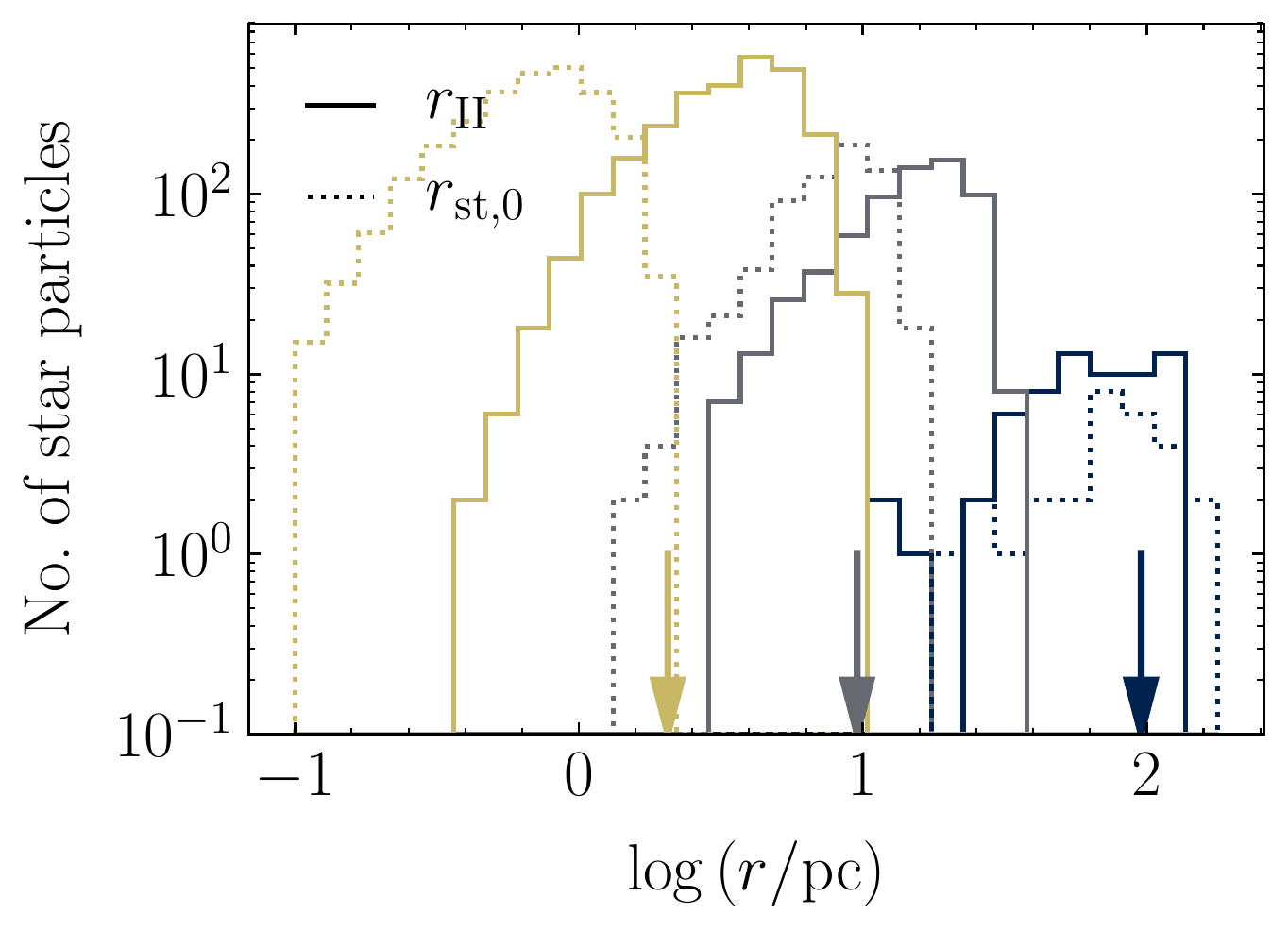}
    \caption{Distributions of ionisation front radii (solid lines) and Str\"{o}mgren radii (dotted lines) for the star particles in our three `HII heat \& beamed mom.' simulations at time $t=600$~Myr. Each line colour corresponds to a different numerical mass resolution: $10^5~{\rm M}_\odot$ (LOW, dark blue), $10^4~{\rm M}_\odot$ (MED, grey), and $10^3~{\rm M}_\odot$ (HI, yellow). The arrows point to the median gas cell radius in each galaxy.}
\end{figure}
% ====================

To obtain the momentum injected into the cloud per unit time, we must solve the equation of motion for the expansion of the shell. We assume that the contribution of $P_{\rm gas}$ is well-approximated by its value in a gas pressure-dominated HII region, for which $P_{\rm gas} \gg P_{\rm rad}$. In the case that $P_{\rm gas} \ll P_{\rm rad}$, the inaccuracy associated with this assumption will be small in comparison to the radiation pressure. Once the rate of expansion $\dot{r}_{\rm II}$ slows to the speed of sound in the ionised gas, the density inside the HII region equilibriates to a uniform value on the sound-crossing time, and is described by the condition of photoionisation balance~\citep{Spitzer78}, such that
\begin{equation}
\label{Eqn::photoionisationbalance}
\frac{4}{3} \pi r_{\rm II}^3 \alpha_{\rm B} \Big(\frac{\rho_{\rm II} c_{\rm II}^2}{k_{\rm B} T_{\rm II}}\Big)^2 \eta^2 = \phi S,
\end{equation}
where $\alpha_{\rm B}$ is the case-B recombination coefficient and $\phi$ is a dimensionless constant that quantifies the effect of photon absorption by dust grains.\footnote{This accounts for the 27~per~cent of photons at Milky Way metallicity~\protect\citep{McKee&Williams97} that are absorbed by dust grains and so do not contribute to the gas pressure. Following KM09, we assume that the gas and dust are well-coupled by the ambient magnetic field, so that the direct radiation pressure does not depend on the gas-to-dust ratio. For further details, see the Appendix of KM09.} The parameter $\eta$ is given by
\begin{equation}
\label{Eqn::eta}
\eta = \frac{\mu}{\sqrt{\mu_e \mu_{\rm H^+}}},
\end{equation}
where $\mu_{\rm H^+}$ is the mean mass per proton in the ionised region, $\mu_{e}$ is the mean mass per free electron, and $\mu$ is the mean mass per free particle. Combining Equations (\ref{Eqn::Pgas}) and (\ref{Eqn::photoionisationbalance}) to rewrite the gas pressure term in Equation (\ref{Eqn::momentumequation_orig}) gives the momentum delivered to the host cloud per unit time as
\begin{align}
\label{Eqn::momentumequation_rdim}
\frac{\dd p}{\dd t} &= \frac{\dd}{\dd t} (M_{\rm sh} \dot{r}_{\rm II}) = \frac{f_{\rm trap} \epsilon_0 S}{(2,1)c} \Big[1 + x_{\rm II}^{\tfrac{1}{2}}\Big] \\
&\approx (1.2, 2.4) \times 10^3 \: S_{49} \: M_\odot \: {\rm km} \: {\rm s}^{-1} {\rm Myr}^{-1} \Big[1 + x_{\rm II}^{\tfrac{1}{2}}\Big],
\end{align}
where we define the dimensionless scale parameter as $x_{\rm II} = r_{\rm II}/r_{\rm ch}$ with
\begin{align}
\label{Eqn::r_ch}
r_{\rm ch} &= \frac{\alpha_{\rm B}}{12(4,1)\pi \phi} \Big(\frac{\epsilon_0 \eta}{k_{\rm B} T_{\rm II}}\Big)^2 \Big(\frac{f_{\rm trap}}{c}\Big)^2 S \\
&\approx (0.5, 1.9) \times 10^{-2} S_{49} \; {\rm pc},
\end{align}
which is the characteristic radius at which the gas and radiation pressure make equal contributions to the rate of momentum injection. To obtain the order-of-magnitude estimates for $\dd p/\dd t$ and $r_{\rm ch}$ in Equations (\ref{Eqn::momentumequation_rdim}) and (\ref{Eqn::r_ch}), we have used the same fiducial values as in KM09, setting $T_{\rm II} = 7000$~K, $\alpha_{\rm B} = 3.46 \times 10^{-13}$~cm$^3$~s$^{-1}$ and $\phi = 0.73$, consistent with the Milky-Way dust-to-gas ratio~\citep{McKee&Williams97}. We set $f_{\rm trap} = 8$, consistent with observations of the pressure inside young HII regions by~\cite{2021ApJ...908...68O}. We take $\eta = 0.48$\footnote{The condition for photo-ionisation balance given in Equation (\ref{Eqn::photoionisationbalance}) differs from Equation (2) of KM09 by a factor of $\eta^2 \sim 0.2$, and therefore the value of $r_{\rm ch}$ that we obtain is smaller than theirs by the same factor.}, corresponding to a ten-to-one ratio of hydrogen to helium atoms in the neutral gas, with the helium atoms singly-ionised. The ionising luminosity has been rescaled as $S_{49} = S/10^{49} {\rm s}^{-1}$. We assume that the volume density of the gas swept up by the HII region is approximately uniform, so that $k_\rho = 0$. The time-evolution of $\dd p/\dd t$ can be computed by writing Equation (\ref{Eqn::momentumequation_rdim}) in the following non-dimensional form
\begin{equation}
\label{Eqn::momentumequation_nondim}
\frac{\dd}{\dd \tau}\Big(x_{\rm II}^{\rm 3-k_\rho} \frac{\dd}{\dd \tau} x_{\rm II}\Big) = 1 + x_{\rm II}^{1/2},
\end{equation}
where $\tau = t/t_{\rm ch}$ for a characteristic time $t_{\rm ch}$ at which the gas and radiation pressure are equal, given by
\begin{equation}
\begin{split}
\label{Eqn::t_ch}
t_{\rm ch} &= \sqrt{\frac{4\pi c \overline{\rho}(r_{\rm st,0})}{3 f_{\rm trap} \epsilon_0 S} r_{\rm ch}^4 \Big(\frac{r_{\rm ch}}{r_{\rm st,0}}\Big)^{-k_\rho}} \\
&\approx (45,333) \: \overline{n}_{\rm H,2}^{1/6} S_{49}^{7/6} \: {\rm yr}
\end{split}
\end{equation}
where $r_{\rm st,0}$ is the initial Str{\"o}mgren radius and $\overline{\rho}(r_{\rm st,0})$ is the mean density inside $r_{\rm st,0}$ in the initial molecular cloud. These two quantities are related by Equation (\ref{Eqn::photoionisationbalance}) as
\begin{equation}
\begin{split}
\label{Eqn::r_st0}
r_{\rm st,0} &= \Big(\frac{3\phi S}{4\pi \alpha_{\rm B}}\Big)^{1/3} \Big(\frac{\mu m_{\rm H}}{\overline{\rho}(r_{\rm st,0})\eta}\Big)^{2/3} \\
&\approx 2.5 \: {\rm pc} \: S_{49}^{1/3} \overline{n}_{\rm H,2}^{-2/3}.
\end{split}
\end{equation}
To obtain Equation (\ref{Eqn::t_ch}), we have used the radial scaling of the mean volume density to write $M_{\rm sh} = (2,4) \pi r_{\rm II}^3 \overline{\rho}(r_{\rm st,0}) (r_{\rm II}/r_{\rm st,0})^{-k_\rho}$. The numerical value of $t_{\rm ch}$ is obtained by writing $\overline{\rho}_{\rm st,0} = 100 \overline{\mu} m_{\rm H} \overline{n}_{\rm H,2}$ as in KM09, where $\overline{\mu} \sim 1.4$ is the mean molecular weight in the ionised gas, $m_{\rm H}$ is the proton mass, and $\overline{n}_{\rm H,2}$ is the number density of hydrogen atoms inside $r_{\rm st, 0}$, in units of $100$~cm$^{-3}$.

Equation (\ref{Eqn::momentumequation_nondim}) has an approximate analytic solution that interpolates between the gas-dominated and radiation-dominated cases to an accuracy of better than $5$~per~cent~\citep{KrumholzMatzner09}, given by
\begin{equation}
\label{Eqn::xapprox}
x_{\rm II, approx} = \Big[\frac{3}{2}\tau^2 + \Big(\frac{25}{28} \tau^2\Big)^{6/5}\Big]^{1/3}.
\end{equation}
With this solution, we may finally write the momentum equation as
\begin{equation}
\begin{split}
\label{Eqn::momentumeqn_final}
\frac{\dd p}{\dd t} \approx (1.2, 2.4) \: &S_{49} \times 10^3 M_\odot \: {\rm km} \: {\rm s}^{-1} {\rm Myr}^{-1} \times \\
&\left\{ 1 + \Big[\frac{3}{2}\frac{t^2}{t_{\rm ch}^2} + \Big(\frac{25}{28} \frac{t^2}{t_{\rm ch}^2}\Big)^{6/5}\Big]^{1/6}\right\},
\end{split}
\end{equation}
with $t_{\rm ch}$ given by Equation (\ref{Eqn::t_ch}). An HII region will deposit momentum at this rate into the surrounding dense gas until its expansion stalls (see Section~\ref{Sec::stalling}). For a Galactic giant molecular cloud with $\overline{n}_{\rm H,2} \sim 1$ and a star cluster with an ionising luminosity of $S_{49} \sim 100$, this characteristic time is around $10,000$ years, indicating that for the majority of its life-span (of order a few Myr), the momentum output from such an HII region is dominated by gas pressure. The gas pressure contributes over 90~per~cent of the final injected momentum. Only for the most luminous clusters (such as M82 L$^{\rm a}$ in KM09, with $t_{\rm ch} \sim 10$~Myr) does radiation pressure dominate the momentum budget. We note that other sub-grid models for HII region feedback at similar resolutions~\citep[e.g.][]{2011MNRAS.417..950H,2013MNRAS.434.3142A,Agertz13,2014MNRAS.445..581H,2015ApJ...804...18A} consider only the part of $\dd p/\dd t$ due to radiation pressure (the first term in the curly brackets in Equation~\ref{Eqn::momentumeqn_final}). In order to inject significant quantities of momentum from HII regions, they therefore inflate $f_{\rm trap}$ above observed values. This is discussed further in Section~\ref{Sec::SNe-fb}.

\subsection{Numerical implementation of HII region momentum} \label{Sec::num-methods}
\subsubsection{Grouping of star particles} \label{Sec::grouping}
\begin{figure}
  \label{Fig::FoF-schematic}
    \includegraphics[width=\linewidth]{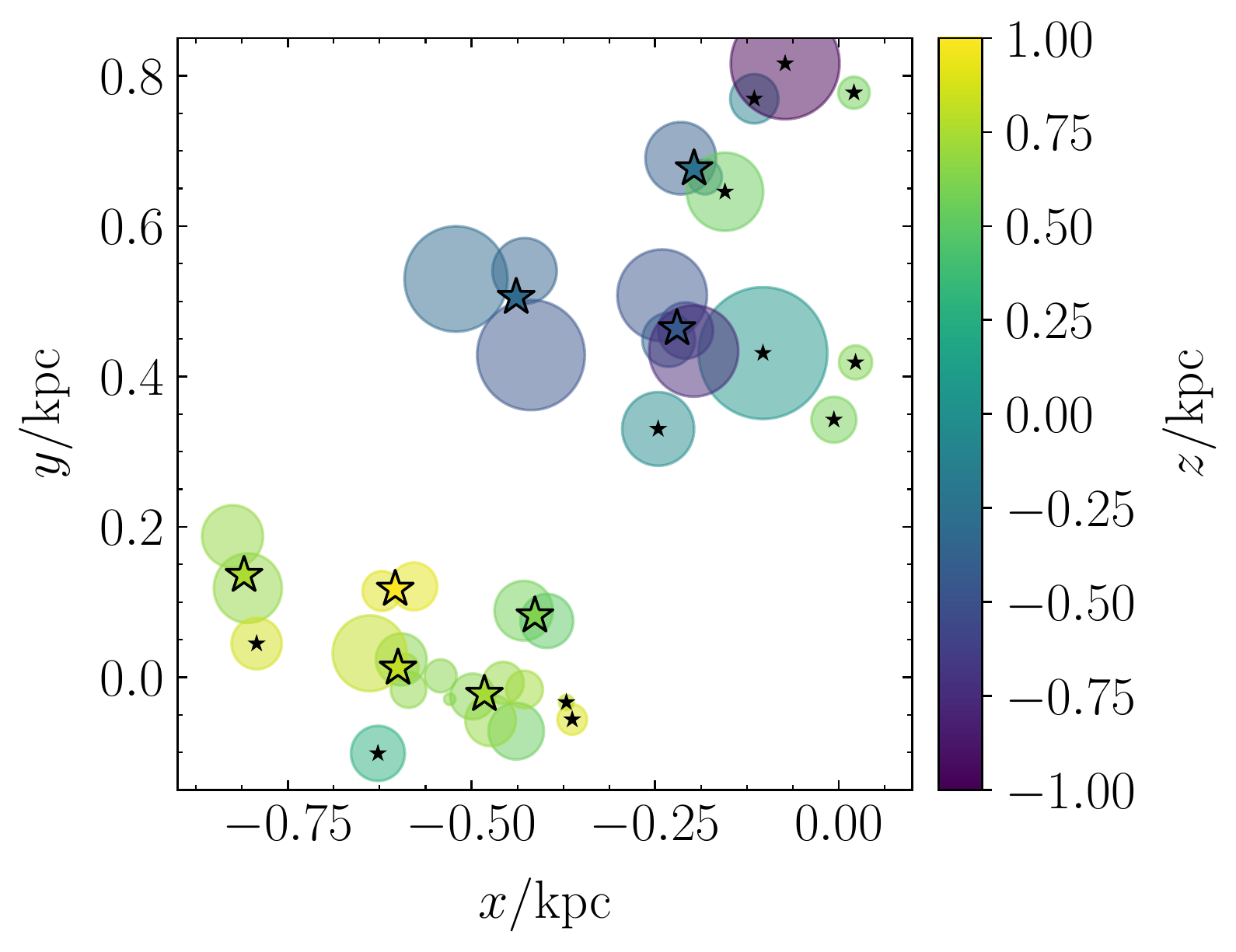}
    \caption{FoF groups of HII regions with overlapping ionisation front radii in a $(950~{\rm pc})^3$ box within the LOW-resolution simulation with spherical HII region feedback (`HII spherical mom.') at $\sim 600~{\rm Myr}$ (the galaxy is centred at the origin). The coloured circles represent the ionisation front radii of individual HII regions, while the small black stars mark the positions of unlinked HII regions (FoF group size of one) and the coloured stars mark the luminosity-weighted centres of the linked HII regions (FoF group size greater than one). The colour bar gives the z-position of the HII regions.}
\end{figure}

The rate of momentum injection given in Equation~(\ref{Eqn::momentumeqn_final}) does not scale linearly with the cluster luminosity $S_{49}$. This means that the total momentum injected by the star particles in a numerical simulation will not trivially converge with increasing mass resolution. As shown in the left-hand panel of Figure~\ref{Fig::FoF-convergence-test}, the maximum stellar particle mass in {\sc Arepo} is equal to twice the simulation mass resolution (the median gas cell mass, given by the solid vertical lines). Upon reaching the star formation threshold $\rho_{\rm thresh}$, larger gas cells are decremented in mass and `spawn' star particles at twice the simulation resolution, while smaller gas cells are deleted and replaced by stars of equal mass. At mass resolutions of $\sim 900~{\rm M}_\odot$, the largest stellar clusters are made up of hundreds of star particles with overlapping ionisation-front radii $r_{\rm II,*}$, each of which is given by
\begin{equation}
\begin{split}
\label{Eqn::rII}
r_{{\rm II}, *}(t) &= r_{\rm ch} x_{\rm II, approx} \\
&\approx 0.5 \times 10^{-2} S_{49,*} \: {\rm pc} \: \times \\
&\left\{\frac{3}{2} \Big(\frac{t_*}{t_{{\rm ch},*}}\Big)^2 + \Big[\frac{25}{28} \Big(\frac{t_*}{t_{{\rm ch},*}}\Big)^2\Big]^{6/5}\right\}^{1/3},
\end{split}
\end{equation}
where $t_{\rm ch,*}$ is the characteristic time for an individual star particle, (see Equation~\ref{Eqn::t_ch}). Physically, a group of star particles whose ionisation fronts overlap should be treated as a single HII region with a single birth density $\overline{n}_{\rm H,2}$ and luminosity $S_{49}$, given that the density of the ionised gas inside the bounding shell of a subsonic HII region equilibriates on its sound-crossing time and becomes uniform. We therefore substantially improve the resolution-convergence of our momentum deposition (in a physically-motivated sense) by using a Friends-of-Friends (FoF) linking algorithm between star particles, with a linking length of $r_{\rm II, *}$.\footnote{Note that this method cannot produce perfect resolution convergence because the ionisation front radii also depend weakly on both the stellar luminosity $S_{49}$ and the stellar birth density $\overline{n}_{\rm H,2}$.} A sample of the FoF groups produced by this algorithm in the low-resolution run is shown in Figure~\ref{Fig::FoF-schematic}.

In practice, the total momentum injected by an FoF-grouped HII region during a numerical time-step $\Delta t_{\rm HII}$ is then given by
\begin{equation}
\Delta p_{\rm HII} = {\Big(\frac{\dd p}{\dd t}\Big)_{\rm FoF} \Delta t_{\rm HII}},
\end{equation}
with a momentum injection rate of
\begin{equation}
\begin{split}
\label{Eqn::momentumequation_fof}
\Big( \frac{\dd p}{\dd t} &\Big)_{\rm FoF} = \; (1.2, 2.4) \times 10^3 \times \\ &\sum_{*=1}^N{S_{49,*}} \: {\rm M}_\odot \: {\rm km} \: {\rm s}^{-1} {\rm Myr}^{-1} \times \\
 &\left\{1 + \Big[\frac{3}{2} \Big(\frac{\langle t \rangle_S}{t_{\rm ch, FoF}}\Big)^2 + \Big(\frac{25}{28} \frac{\langle t \rangle_S}{t_{\rm ch,FoF}}\Big)^{6/5}\Big]^{1/6} \right\},
\end{split}
\end{equation}
and a characteristic time of
\begin{equation}
\label{Eqn::t_ch_fof}
t_{\rm ch,FoF} = \sqrt{(0.6, 33) \Big(\sum^N_{*=1}{S_{49,*}}\Big)^{7/3} \langle \overline{n}_{\rm H,2} \rangle_S^{1/3} \: {\rm pc} \: {\rm s}^2}
\end{equation}
for each FoF group. The angled brackets $\langle ... \rangle_S$ denote ionising luminosity-weighted averages over the star particles $*=1 ... N$ in the group, such that $\langle t \rangle_S$ is the luminosity-averaged age of the star particles. The momentum is injected at the luminosity-weighted centre of the group, given by
\begin{equation}
\label{Eqn::centre_fof}
\langle \bm{x} \rangle_S = \frac{\sum_{*=1}^N{S_{49,*}\bm{x}_*}}{\sum_{*=1}^N{S_{49,*}}}.
\end{equation}
To ensure that all star particles in a single group have their ionising luminosities, ionisation front radii and ages updated on the same time-step, we set $\Delta t_{\rm HII}$ to be the global time-step for the simulation. This means that we inject HII region feedback on global time-steps only, which have a maximum value of $0.1$~Myr for the simulations presented in this work.\footnote{Due to the hierarchical time-stepping procedure used in {\sc Arepo}~\citep[see][for details]{Springel10}, the ionising properties of different star particles from the same FoF group would otherwise be updated at different time intervals. Our global time-step has a maximum value of $0.1$~Myr, and the peak ionisation-front expansion rate for the most massive star particles in our high-resolution simulation ($\sim 2 \times 10^3$~M$_\odot$, see Figure~\ref{Fig::FoF-convergence-test}) is $\sim 20$~pc Myr$^{-1}$, so in the very worst case, our FoF groups may be affected by an error of order $\sim 2$~pc.}

% ==================== GROUPING SCHEMATIC
% ====================

%========================= 2HII CONVERGENCE FIG
\begin{figure}
  \label{Fig::2HII-convergence}
    \includegraphics[width=\linewidth]{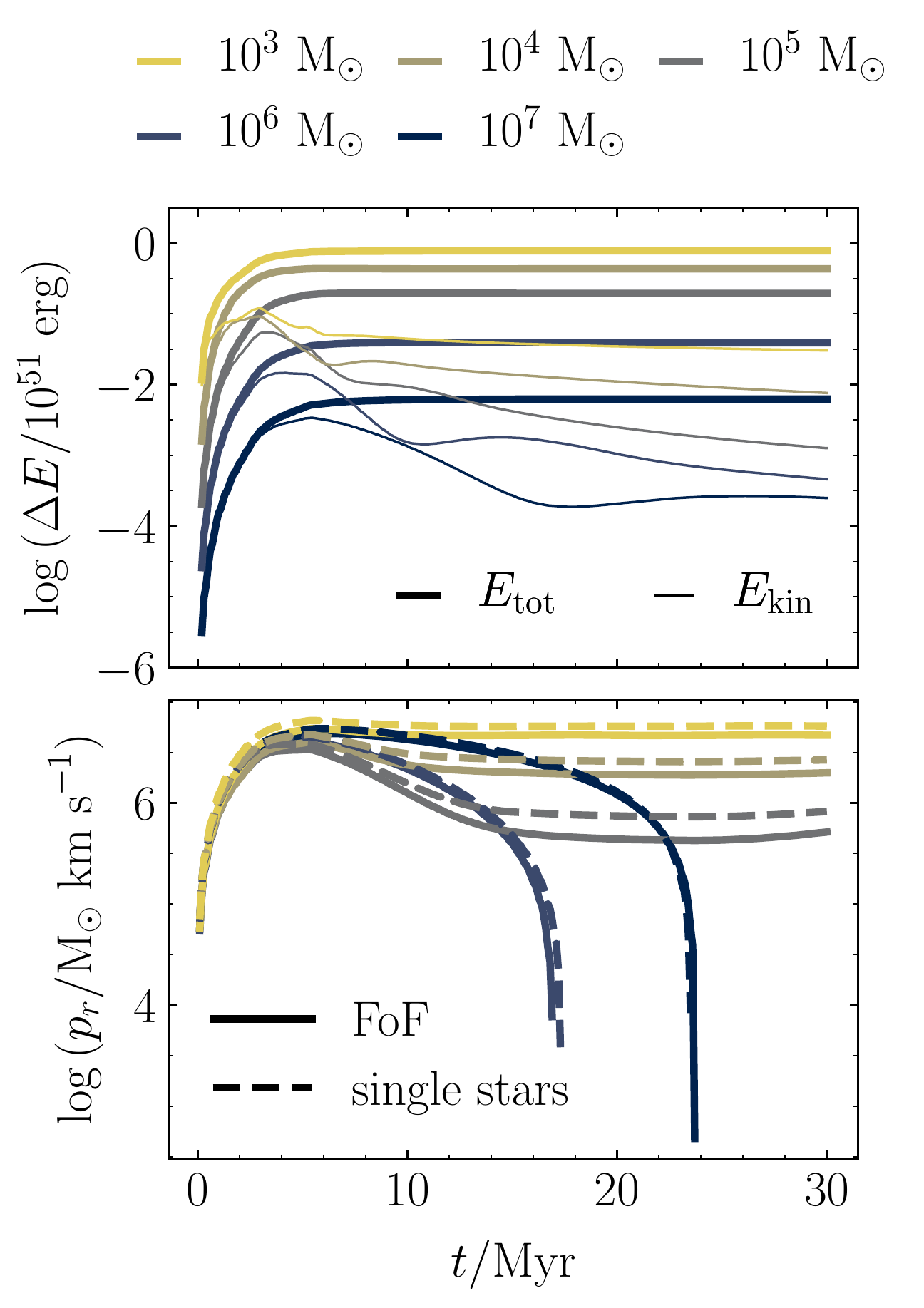}
    \caption{Energy (upper panel) and radial momentum (lower panel) of the gas cells in a box of size $(950~{\rm pc})^3$ and density $100~{\rm cm}^3$ containing two star particles of mass $1 \times 10^4~{\rm M}_\odot$, separated by a distance of $5$~pc, as a function of time. The star particles both inject spherical momentum from HII regions (no thermal energy injection) according to the prescription outlined in Section~\ref{Sec::num-methods}. In the lower panel, the combined momentum from the individual star particles (dashed lines) is compared to the momentum injected by the FoF-grouped pair (solid lines). The resolution varies from a lowest value of $10^7~{\rm M}_\odot$ per gas cell (dark blue) up to a highest resolution of $10^3~{\rm M}_\odot$ per gas cell (yellow).}
\end{figure}
%=========================

%========================= BLISTER PROJECTIONS
\begin{figure*}
  \label{Fig::blister-projections}
    \includegraphics[width=\linewidth]{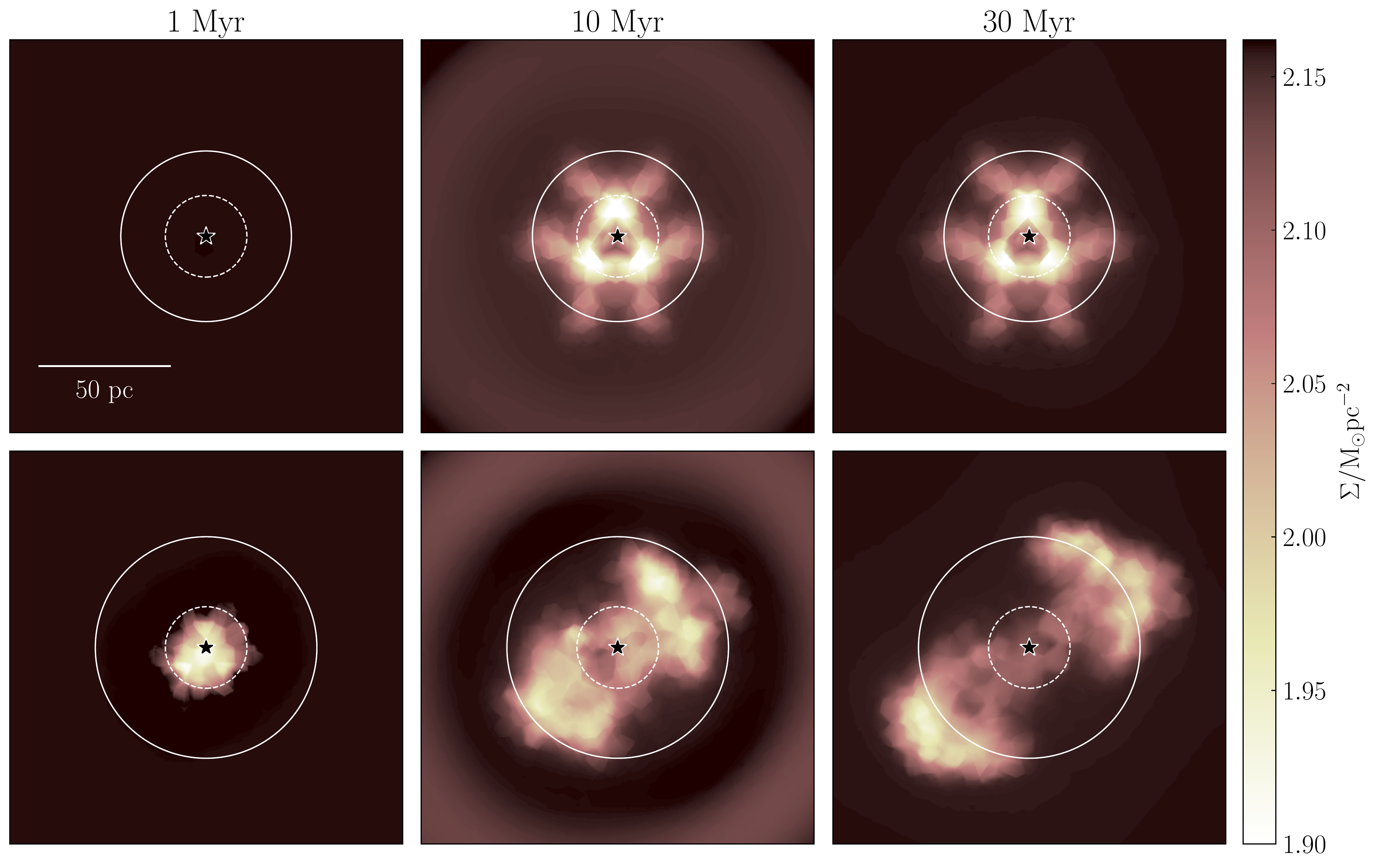}
    \vspace{-0.5cm} \caption{Column density projections of the gas cells surrounding an HII region of stellar mass $10^4$~M$_\odot$ in a $(950~{\rm pc})^3$ box containing $128^3$ gas cells at a volume density of $100~{\rm cm}^{-3}$. The top row shows a spherical/embedded HII region, while the bottom row shows a beamed/blister-type HII region with momentum injected accordering to Equation (\ref{Eqn::jet-profile}), using an opening angle of $\Theta = \pi/12$ and axis aligned along $(x=1/\sqrt{2}, y=1/\sqrt{2})$. All projections are computed parallel to the simulation $z$-axis.}
\end{figure*}
%=========================

In Figure~\ref{Fig::FoF-convergence-test} we show the effect of grouping on the HII region masses (left-hand panel), the ionising luminosities (centre panel), and the momentum injection rate (right-hand panel) in the low-resolution (dark blue lines), medium-resolution (grey lines) and high-resolution (yellow lines) Agora disc simulations. Comparison of the bold lines (FoF groups) and thin lines (star particles) demonstrates that the distributions of stellar birth masses $m_{\rm birth}$, luminosities $S_{49}$, and momentum injection rates $\dd p/\dd t$, are brought closer to convergence by the FoF grouping. The ionisation front radii $r_{\rm II}$ used to compute the groups are displayed for each simulation in Figure~\ref{Fig::rst0}. We might also consider using the Str{\"o}mgren radius $r_{\rm st,0}$ (dotted lines, left-hand panel) as the FoF linking-length, however $r_{\rm st,0}$ is more heavily-dependent on the stellar birth density than is $r_{\rm II}$ (see Equations~\ref{Eqn::r_st0} and~\ref{Eqn::rII}), and so is more heavily-dependent on the simulation resolution, making it a less favourable choice.

While our FoF grouping corrects for the non-linear dependence of Equation~(\ref{Eqn::momentumeqn_final}) on the ionising luminosity, and reduces the spurious cancellation of the momentum injected between adjacent star particles, it \textit{does not} address resolution-dependent variations in the spatial distribution of stellar mass in our simulations, caused partly by the variation in star particle mass, and partly by the suppressed clustering of star particles at lower resolutions. These effects change the spatial distribution of the energy injected by stellar feedback. As discussed in~\cite{2020arXiv200911309S,2020arXiv200403608K}, an increase in the clustering of supernovae leads to burstier feedback with larger outflows perpendicular to the galactic mid-plane. We discuss these effects further in Section~\ref{Sec::discussion}.

%========================= HEAT CONVERGENCE FIG
\begin{figure}
  \label{Fig::heat-convergence}
    \includegraphics[width=\linewidth]{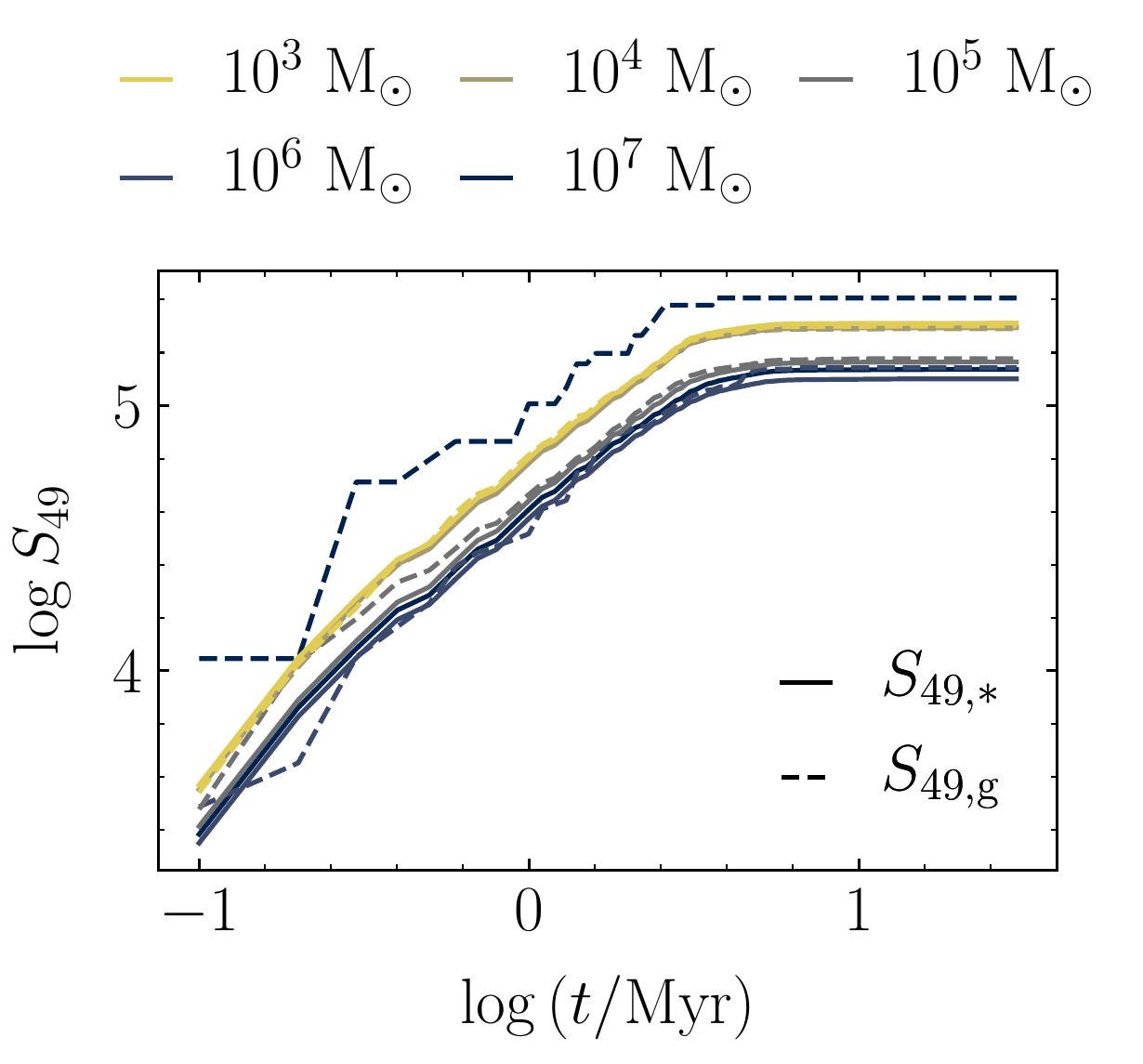}
    \caption{Ionising luminosity emitted by star particles ($S_{49, *}$, solid lines) and absorbed by surrounding gas cells ($S_{49, {\rm g}}$, thin lines) at a density of $100~{\rm cm}^{-3}$ in a box of size $(950~{\rm pc})^3$ containing 100 star particles of mass $1 \times 10^4~{\rm M}_\odot$, as a function of time. The star particles inject HII region heating feedback according to the prescription outlined in Section~\ref{Sec::num-methods}. The resolution varies from a lowest resolution of $10^7~{\rm M}_\odot$ per gas cell (dark blue) up to a highest resolution of $10^3~{\rm M}_\odot$ per gas cell (yellow). Note that lines for the $10^3~{\rm M}_\odot$ and $10^4~{\rm M}_\odot$ overlap.}
\end{figure}
%=========================

%========================= DISC MORPHOLOGY
\begin{figure*}
  \label{Fig::morphology}
    \includegraphics[width=\linewidth]{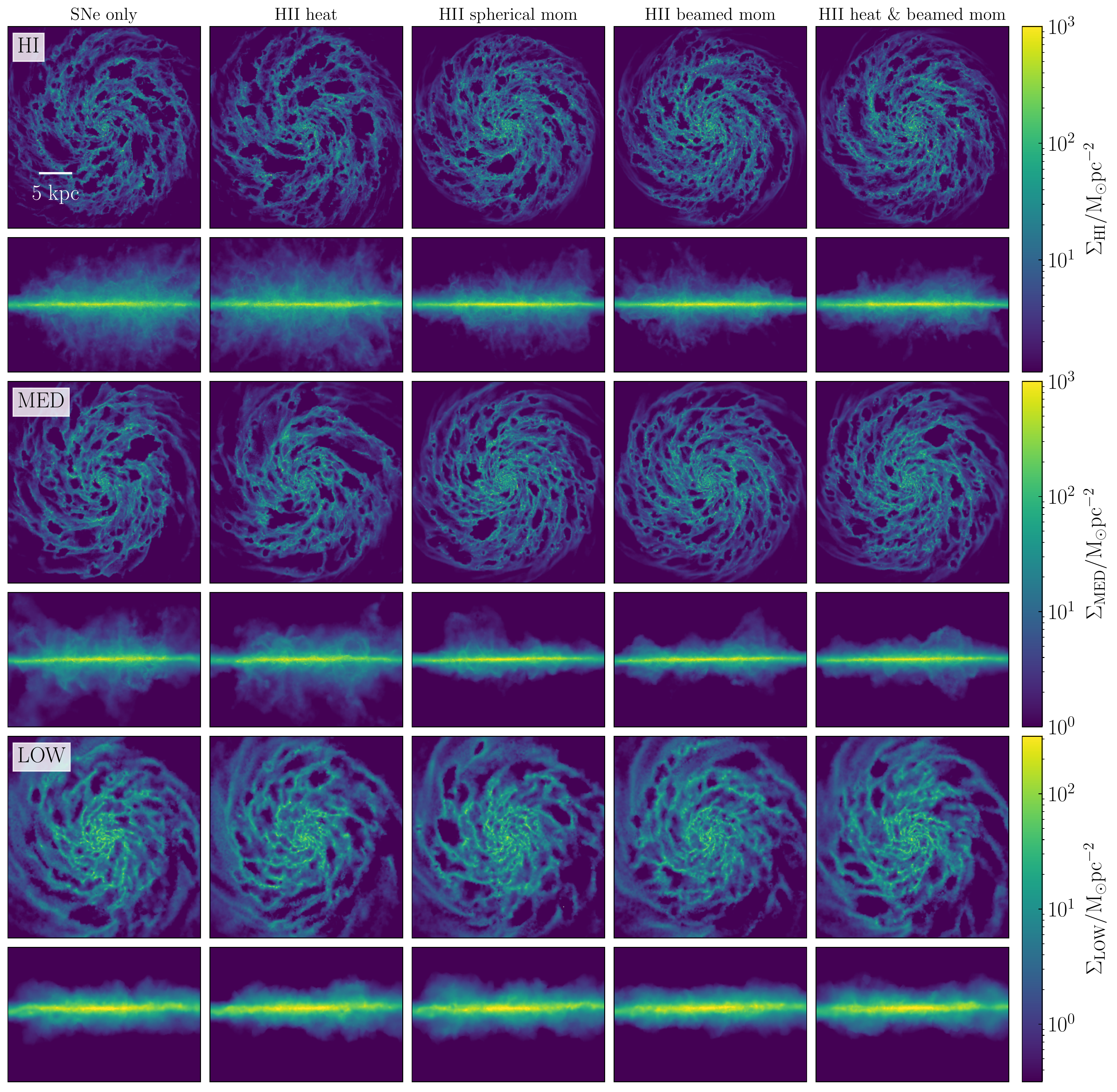}
    \caption{Column density maps of the gas at each resolution (LOW = $10^5~{\rm M}_\odot$ per cell, MED = $10^4~{\rm M}_\odot$, HI = $10^3~{\rm M}_\odot$) and with each feedback prescription, at $t=600$~Myr. Momentum injection from HII regions changes the phase structure of the interstellar medium in the MED- and HI-resolution cases only. It also improves the numerical convergence of the gas-disc scale-height. Heating from HII regions makes no difference to the disc morphology at any resolution.}
\end{figure*}
%=========================

%========================= TUNING FORKS
\begin{figure}
  \label{Fig::tuning-forks}
    \includegraphics[width=\linewidth]{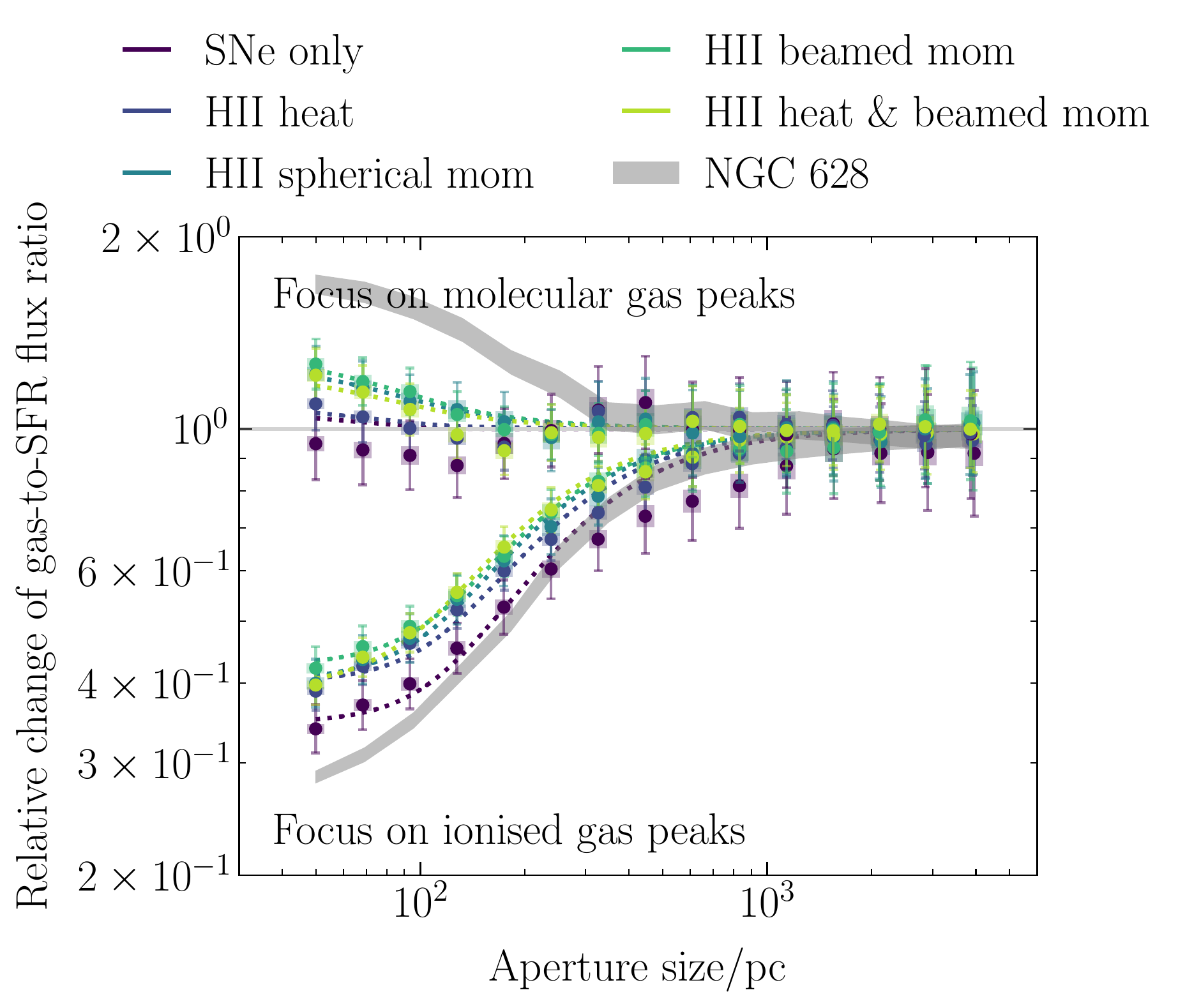}
    \caption{The gas-to-SFR flux ratio relative to the galactic average value as a function of aperture size, for each of our high-resolution galaxies at $t = 600$~Myr. The upper branch represents apertures focussed on molecular gas peaks, while the lower branch represents apertures focussed on the peaks of surface density of `young stars' (ages $0$-$5$~Myr). The error bars on each data point represent the 1$\sigma$ uncertainty on the value of the gas-to-SFR flux ratio. The shaded areas on each data point indicate the effective $1\sigma$ uncertainty range that accounts for the covariance between the data points. The grey-shaded region shows the result of applying the same analysis to observations of the Milky Way-like galaxy NGC 628~\protect\citep{Chevance20}. The dashed coloured lines represent the best-fit model of~\protect\cite{Kruijssen18a}.}
\end{figure}
%=========================

%========================= SFHs and OUTFLOWS
\begin{figure*}
  \label{Fig::outflows}
    \includegraphics[width=\linewidth]{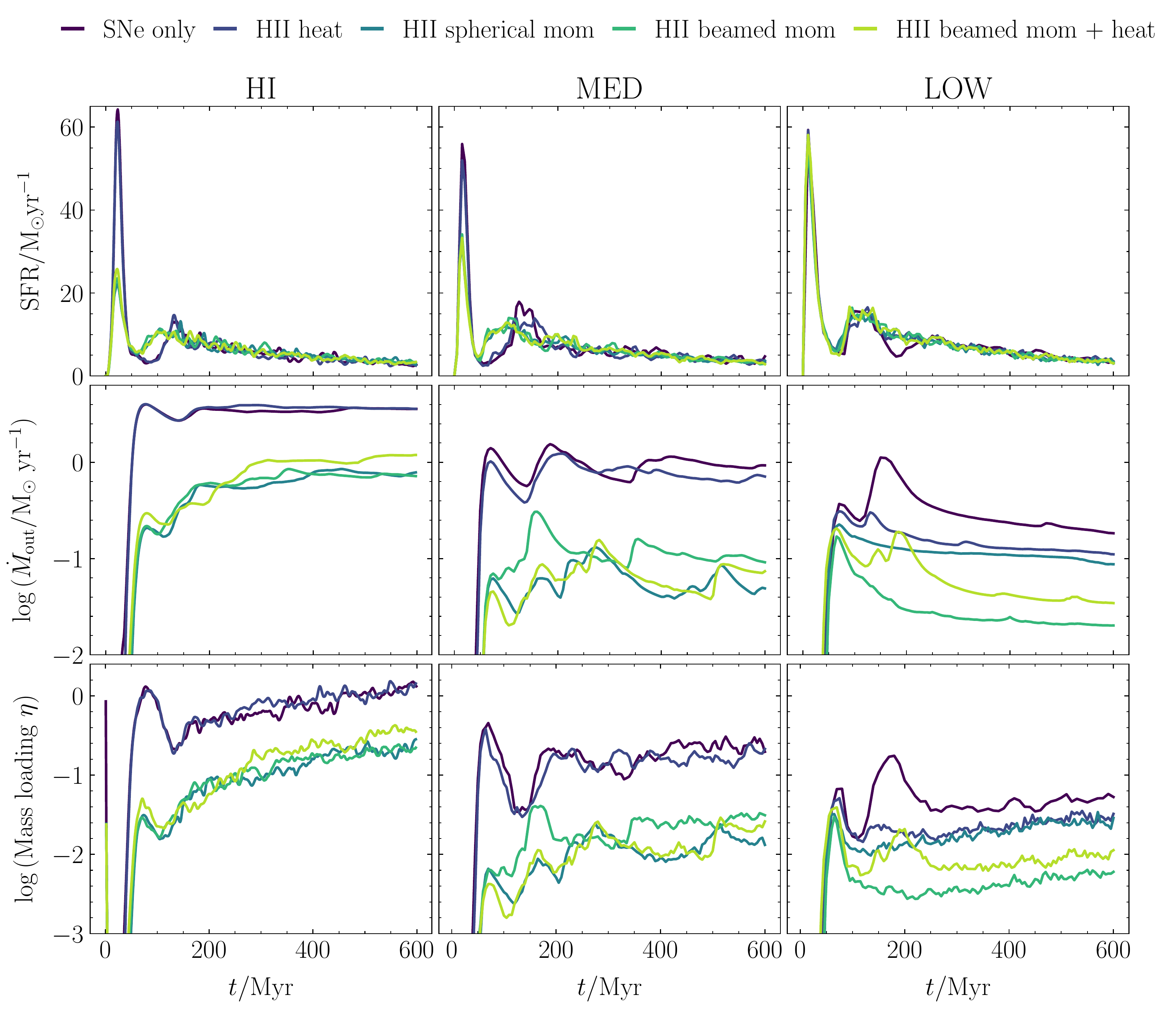}
    \caption{Global galactic star formation rate (top row), gas outflow rate (see Section~\protect\ref{Sec::outflows}) from the galactic mid-plane (middle row) and mass-loading of outflows (bottom row) as a function of simulation time for the simulation with both thermal and beamed momentum from HII regions (HII thermal \& beamed mom.), at low-resolution (LOW, dark blue), at medium-resolution (MED, grey), and high-resolution (HI, yellow).}
\end{figure*}
%=========================

%========================= VSS
\begin{figure*}
  \label{Fig::vss}
    \includegraphics[width=\linewidth]{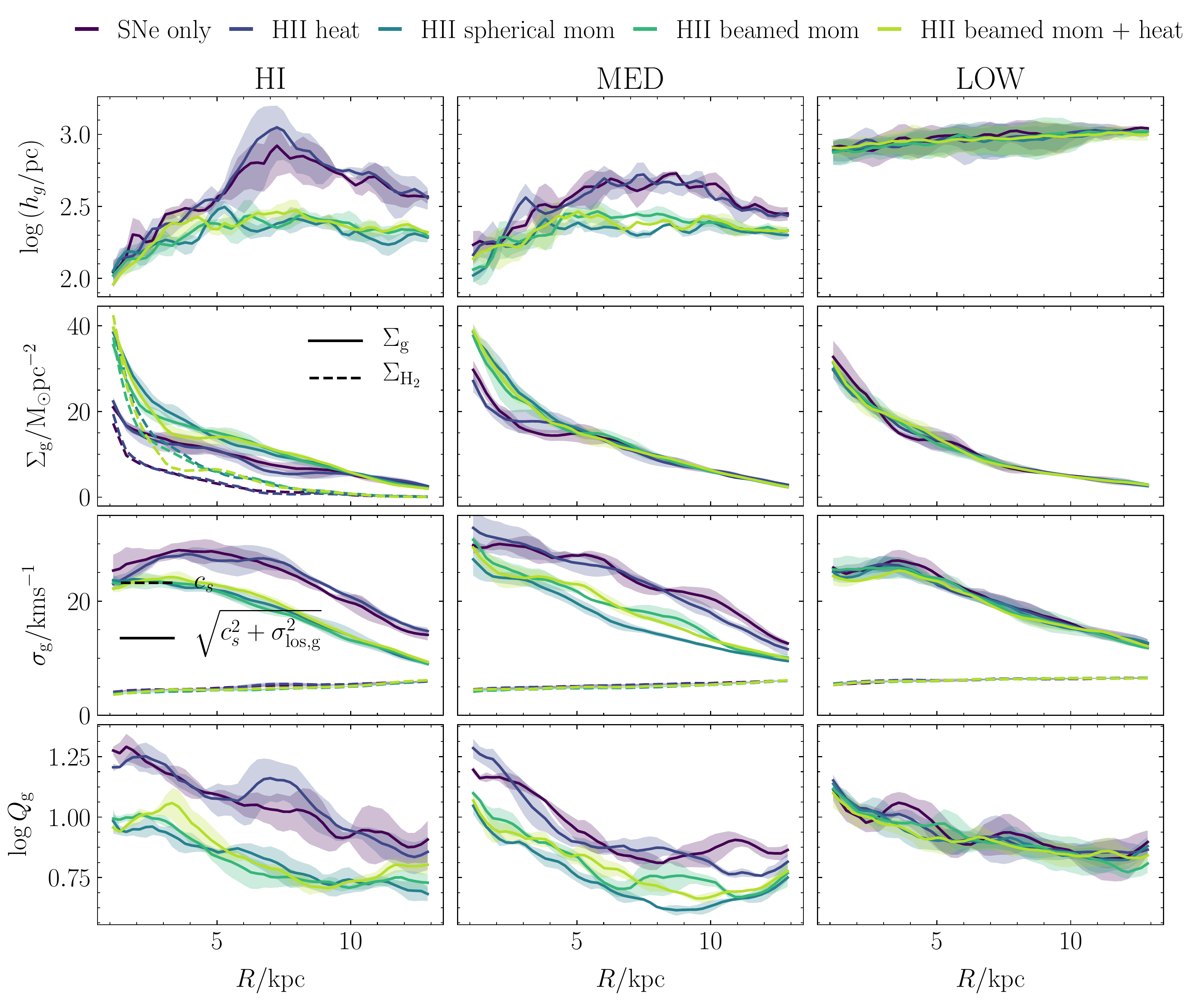}
    \caption{Azimuthally- and temporally-averaged cold-gas ($\leq 10^4~{\rm K}$) gas-disc scale-height (top row), molecular gas (dashed lines) and total gas (solid lines) surface density (second row), thermal (dashed lines) and total (thermal plus turbulent, solid lines) velocity dispersion (third row), and the Toomre $Q_{\rm g}$ parameter of the total gas distribution (bottom row) as a function of the galactocentric radius for each simulated galaxy at each resolution, across the time interval $t=300$-$600$~Myr.}
\end{figure*}
%=========================

\subsubsection{Injection of momentum from HII regions} \label{Sec::injection}
We inject the radial momentum $\Delta p_{{\rm HII}, *}$ from each star particle at the luminosity-weighted centre $\langle \bm{x} \rangle_S$ of its FoF group via the same procedure as used for supernovae, described in~\cite{2020arXiv200403608K,2020MNRAS.498..385J}. Briefly, the algorithm proceeds as follows.
\begin{enumerate}
  \item For each FoF group, find the nearest-neighbour gas particle $j$ to the luminosity-weighted centre of mass.
  \item Increment the total radial momentum $\Delta p_{j, {\rm HII}}$ received by cell $j$ from all of the FoF groups it hosts, such that
\begin{equation}
\Delta p_{j, {\rm HII}} = \sum_{{\rm FoF}=1}^N {\Big(\frac{\dd p}{\dd t}\Big)_{\rm FoF}} \Delta t_{\rm HII}.
\end{equation}
  \item For each gas cell $j$ that has received HII-region momentum, find the set of neighbouring gas cells $k$ with which it shares a Voronoi face. Compute the fraction of the radial momentum received by each facing cell according to
\begin{equation}
\label{Eqn::j-to-k-inj}
\Delta \bm{p}_{k, {\rm HII}} = w_k \hat{\bm{r}}_{j \rightarrow k} \Delta p_{j, {\rm HII}},
\end{equation}
where $\hat{\bm{r}}_{j \rightarrow k}$ is the unit vector from the centre of the host cell to the centre of the cell receiving the feedback, and the weight factor $w_k$ is the fractional Voronoi face area $A_{j \rightarrow k}$ shared between these cells, such that
\begin{equation}
\label{Eqn::weight-fn}
w_k = \frac{A_{j \rightarrow k}}{\sum_k{A_{j \rightarrow k}}}.
\end{equation}
Equation (\ref{Eqn::j-to-k-inj}) ensures that the momentum injection is perfectly isotropic, regardless of the distribution over the volumes of the cells $k$.
  \item Ensure conservation of linear momentum by subtracting the sum of the injected momenta $\sum_k{\Delta p_{k, {\rm HII}}}$ from the momentum of the central cell $j$.\footnote{We note that we do not account for a full tensor renormalisation of the injected momentum, as in~\cite{Hopkins18b,Smith2018,2020arXiv200911309S}, for example.}
\end{enumerate}

In Figure~\ref{Fig::2HII-convergence}, we check the numerical convergence of the momentum and energy injected according to the algorithm described above, at mass resolutions varying between $10^7~{\rm M}_\odot$ and $10^3~{\rm M}_\odot$ per gas cell. We take a box of side-length $950~{\rm pc}$ and uniform gas density $100~{\rm cm}^{-3}$, containing a single pair of star particles of mass $10^4~{\rm M}_\odot$ each. We record the radial momentum of the gas cells in the box as a function of time when the stars inject momentum from their individual HII regions (dashed lines, lower panel), and when the stars are grouped via the FoF procedure described in Section~\ref{Sec::grouping} (solid lines, lower panel). We also record the kinetic and total energies of the gas cells in the FoF-grouped case, represented by the thin and bold lines, respectively, in the top panel of Figure~\ref{Fig::2HII-convergence}. The bottom panel demonstrates that the radial momentum injected is converging to within 1.1 dex in momentum per 3 dex in mass resolution in both the FoF-grouped (solid lines) and ungrouped (dashed lines) cases, for the mass resolutions between $10^3~{\rm M}_\odot$ and $10^5~{\rm M}_\odot$ spanned by our isolated disc galaxies. At lower resolutions the injected momentum does not persist, but rather begins to drop steeply after about $10$~Myr of evolution. This is because the ionisation front bounding the HII region is never resolved at mass resolutions of $>10^5~{\rm M}_\odot$, and so the neighbouring gas cells $k$ have a combined mass much larger than that of the swept-up shell. This greatly reduces their final velocities/kinetic energies (shown in the top panel of Figure~\ref{Fig::2HII-convergence}), and so the injected momentum is quickly lost. This behaviour is not inaccurate, as entirely-unresolved feedback processes should not have any impact on the simulated interstellar medium.

\subsubsection{Directional injection for blister-type HII regions}
The weight factor $w_k$ in Section~\ref{Sec::injection} results in isotropic momentum injection, appropriate for embedded HII regions. To mimic the directional outflow from a blister-type HII region along an axis $\hat{\bm{z}}_{\rm FoF}$, we instead weight the momenta $\Delta p_{k, {\rm HII}}$ by the following axisymmetric factor,
\begin{equation}
\label{Eqn::jet-profile}
\begin{split}
w(\theta_k, A_k) &= \frac{A_{j \rightarrow k} f(\theta_k)}{\sum_k{A_{j \rightarrow k} f(\theta_k)}} \\
f(\theta_k) &= \Big[\log{\Big(\frac{2}{\Theta}\Big)(1+\Theta^2-\cos^2{\theta_k})}\Big]^{-1}
\end{split}
\end{equation}
where $\Theta$ controls the width of the beam and $\theta_k$ is the angle between the beam-axis and the unit vector $\hat{\bm{r}}_{j \rightarrow k}$ connecting cells $j$ and $k$, defined by
\begin{equation}
\cos{\theta_k} = \frac{\hat{\bm{r}}_{j \rightarrow k} \cdot \hat{\bm{z}}_{\rm FoF}}{|\hat{\bm{z}}_{\rm FoF}|}.
\end{equation}
The opening angle is set to $\Theta = \pi/12$ in our simulations, and the beam-axis vector $\hat{\bm{z}}_*$ for each star particle is drawn randomly from a uniform distribution over the spherical polar angles about the star's position, $\phi_*$ and $\theta_*$.  This value is fixed throughout the star particle's lifetime, and the beam-axis $\hat{\bm{z}}_{\rm FoF}$ of each FoF group is calculated as a luminosity-weighted average of $\hat{\bm{z}}_*$ across the constituent star particles. In Figure~\ref{Fig::blister-projections} we compare the density profiles for spherical- (top row) and blister-type (bottom row) momentum injection, at simulation times $1$~Myr, $10$~Myr and $30$~Myr after the birth of the stellar cluster in a uniform medium of density $100~{\rm cm}^{-3}$. Qualitatively, the blister-type momentum injection results in a faster and wider ejection of gas away from the cluster centre than does the spherical momentum injection, despite the fact that the ionisation front radius (solid white lines) is only marginally larger. We note that in this uniform-density box, the number of Voronoi cells surrounding the star particle is relatively small, resulting in a deviation from perfect spherical symmetry when the feedback is injected isotropically (top row of Figure~\ref{Fig::blister-projections}, the momentum propagates along rays joining the star particle to the centroids of the neighbouring cells). This effect will be less marked in the highly-overdense star-forming regions of isolated disc galaxies.

\subsection{Stalling of HII regions} \label{Sec::stalling}
In computing the FoF groups via the method presented in Section~\ref{Sec::grouping}, we must be careful to exclude star particles whose ionisation fronts have stalled, and which are no longer depositing significant quantities of momentum into the surrounding gas. Stalling occurs when the rate of HII region expansion becomes comparable to the velocity dispersion of the host cloud, at which point the ionised and neutral gas are able to intermingle and the swept-up shell loses its coherence~\citep{Matzner02}. After this transition, it no longer makes sense to include the stalled HII region in an FoF group of expanding HII regions, as its radius and internal density are no longer well-defined. In particular, we want to avoid the case where such an HII region links together two active HII regions, spuriously shifting the origin of their momentum ejection to a position halfway between the two particles. Before the FoF groups are calculated, we therefore compute the rate of HII region expansion $\dot{r}_{{\rm II},*}$ for each star particle, and if this is found to be smaller than the velocity dispersion of the surrounding gas at the same scale, we flag the particle as `stalled'. Star particles with stalled ionisation fronts are not allowed to be FoF group members, but are still allowed to contribute to HII region feedback with what little remains of their ionising luminosity. Following KM09, we approximate the ambient velocity dispersion by considering a blister-type HII region centred at the origin of a cloud with an average density of $\overline{\rho}(r) = 3/(3-k_\rho) \rho_0 (r/r_0)^{-k_\rho}$ and a virial parameter $\alpha_{\rm vir}$ as measured on the scale of the HII region. This gives a cloud velocity dispersion of
\begin{equation}
\label{Eqn::sigma}
\begin{split}
\sigma_{\rm cl}(r_{\rm II}) &= \sqrt{\frac{\alpha_{\rm vir} G M(<r_{\rm II})}{5r_{\rm II}}} \\
&= \sqrt{\frac{2\pi}{15} \alpha_{\rm vir} G \overline{\rho}(r_{\rm st,0}) r_{\rm ch}^{2-k_\rho} r_{\rm st,0}^{k_\rho} x_{\rm II}^{2-k_\rho}},
\end{split}
\end{equation}
where we assume the cloud is in approximate virial balance with $\alpha_{\rm vir}=1$, and we again take $k_{\rho} = 0$. Equations (\ref{Eqn::r_ch}), (\ref{Eqn::r_st0}) and (\ref{Eqn::sigma}) then imply that the radius at which expansion stalls is given by
\begin{equation}
\label{Eqn::stalling_velocity}
\begin{split}
\dot{r_{\rm II}} &= \frac{r_{\rm ch}}{t_{\rm ch}} \frac{\dd x_{\rm II}}{\dd \tau} \\
&= 8.8 \times 10^{-3} \: {\rm km} \: {\rm s}^{-1} S_{49} \: \overline{n}_{\rm H,2}^{-1/6} \: {\rm cm}^{1/2} x_{\rm II}^{1/2}.
\end{split}
\end{equation}
We calculate the value of $\dot{r}_{{\rm II},*}=\Delta r_{{\rm II}, *}/\Delta t_*$ numerically for each star particle, where $\Delta r_{{\rm II},*}$ is the increment in the ionisation front radius during the particle's time-step $\Delta t_*$. When $\dot{r}_{{\rm II},*}$ crosses $\sigma_{\rm cl}(r_{{\rm II},*})$, the HII region is considered to have stalled.

\subsection{Heating from HII regions} \label{Sec::HII-heat}
To examine the influence of momentum injection relative to thermal energy injection for the HII region feedback, we run one simulation with thermal HII region feedback (without HII region momentum, `HII thermal'), and one simulation with both thermal and kinetic HII region feedback (`HII thermal + beamed mom.'). In these simulations, we inject thermal energy associated with photo-ionisation heating up to a temperature of $7000$~K. To do this, we piggy-back on the injection procedure for the HII region momentum, depositing thermal energy into the nearest-neighbour gas cell and its immediate neighbours. We do not need to access gas cells beyond these immediate neighbours because the Str{\"o}mgren radii of the FoF groups in our simulations are at best marginally-resolved. We also do not need to deal with overlapping Str{\"o}mgren spheres once the FoF groups have already been computed. We therefore proceed as follows:
\begin{enumerate}
	\item For each FoF group, find the nearest-neighbour gas particle $j$ for its luminosity-weighted centre of mass.
	\item Increment the total number of photons $S_{\rm in}$ per unit time received by this gas cell from all the FoF group centres it hosts, such that the final value is $S_{j, {\rm in}} = \sum_{{\rm FoF} = 1}^N{S_{\rm FoF}}$, where $S_{\rm FoF}$ is the total ionising luminosity of all stars in the FoF group, and $N$ is the number of FoF group centres hosted by $j$.
	\item Compute the number of photons $S_{j, {\rm cons}} = \alpha_B N_{j, {\rm H}} n_{j, e}$ per unit time that can be consumed via the ionisation of the material in gas cell $j$, where $N_{j, {\rm H}}$ is the number of hydrogen atoms in the cell and $n_{j, e}$ is the number density of electrons.
	\item If $S_{j, {\rm in}} < S_{j, {\rm cons}}$, then ionise cell $j$ with a probability $S_{j, {\rm in}}/S_{j, {\rm cons}}$. This ensures that over a large number of gas cells, the number of injected photons converges to $S_{j, {\rm in}}$.
	\item If $S_{j, {\rm in}} > S_{j, {\rm cons}}$, ionise cell $j$ and compute the `residual' ionisation rate $S_{j, {\rm res}}$ to be spread to the facing cells $k$, such that $S_{j, {\rm res}} = S_{j, {\rm in}} - S_{j, {\rm cons}}$. Each ionised cell is heated to a temperature of $7000$~K.
\end{enumerate}
In the case that $S_{j, {\rm in}} < S_{j, {\rm cons}}$, the algorithm ends here. Otherwise we continue as follows.
\begin{enumerate}
	\setcounter{enumi}{6}
	\item For each gas cell $j$ with $S_{j, {\rm res}} > 0$, find the set of neighbouring cells $k$ with which it shares a Voronoi face. Compute the fraction of photons it receives according to
\begin{equation}
S_{k, {\rm in}} = w_k S_{j, {\rm res}},
\end{equation}
where $w_k = A_{j \rightarrow k}/\sum_k{A_{j \rightarrow k}}$, as for the injection of HII region momentum in Section~\ref{Sec::injection}.
	\item Ionise each facing cell $k$ with a probability of $S_{k, {\rm in}}/S_{k, {\rm cons}}$. Summed over the set of facing cells for many HII regions, this ensures that the expectation value for the rate of ionisation converges to $S_{j, {\rm res}}$.
\end{enumerate}
Subsequent to the above procedure for thermal energy injection, the chemistry and cooling for each gas cell is computed using {\sc SGChem}, as described in Section~\ref{Sec::SNe-only}. During this computation, we impose a temperature floor of $7000$~K, which is enforced until the next HII-region update. We rely on the chemical network to collisionally-ionise the gas cells in a manner that is self-consistent with their temperatures. This will only produce an ionisation fraction of $10^{-5}$ when cold gas is heated to $7000$~K, but in the non-equilibrium case, whereby gas cools from much higher temperatures to a floor of $7000$~K, much higher ionisation fractions can be achieved. After the chemistry computation, the ionised cells are unflagged and are ready to absorb more photons.

In Figure~\ref{Fig::heat-convergence}, we check that at mass resolutions between $10^3$ and $10^7~{\rm M}_\odot$, the above method ensures convergence of the quantity of photoionised gas. We consider a box of side-length $950~{\rm pc}$ containing a gas of uniform density $100~{\rm cm}^{-3}$, along with $100$ star particles of mass $10^4~{\rm M}_\odot$ each. We record the total cumulative value of $S_{49, *}$ emitted by these particles as a function of time (solid lines), as well as the total cumulative $S_{49, {\rm g}}$ absorbed by the surrounding gas cells (dashed lines). Cooling and chemistry are switched on. We see that the bold and solid lines match at all resolutions, indicating that none of the emitted photons are `wasted' by our restriction of photon injection to the set of facing cells $k$ surrounding each star particle. The offset for the lowest-resolution ($10^7~{\rm M}_\odot$ per gas cell) case is due to the stochastic procedure for choosing the gas cells to ionise: in the limit of a very large number of HII regions ($\gg 100$), we would expect this offset to approach zero. The star particle mass used in this test is in the 99th percentile for FoF grops in the highest-resolution isolated disc simulation used in this work ($10^3~{\rm M}_\odot$ per gas cell), and the gas density is ten times lower than the birth density of these star particles. If all photons are absorbed in this case, then the algorithm described above is valid in its modelling of the heating due to the vast majority of our marginally-resolved HII regions.

\section{Results} \label{Sec::results}
In this section, we analyse the properties of the four simulated disc galaxies with thermal HII region feedback (`HII heat'), spherically-injected HII region momentum (`HII spherical mom.'), blister-type HII region momentum with $\Theta = \pi/12$ (`HII beamed mom.'), and a combination of blister-type momentum and thermal energy (`HII heat \& beamed mom.'), relative to our control simulation with supernova feedback on its own (`SNe only'). The simulations are summarised in Table~\ref{Tab::sims}. We consider the morphology, stability, global star formation rate and phase structure of the interstellar medium (Section~\ref{Sec::disc-props}), and the distribution of the lifetimes, masses, star formation rate densities, and velocity dispersions of its molecular clouds (Section~\ref{Sec::molecular-props}).

In this section, whenever we compare to observed quantities involving molecular hydrogen, we use synthetic $^{12}{\rm CO}(J=1\rightarrow 0)$ maps obtained by post-processing the simulations using the {\sc Despotic} code~\citep{Krumholz14}, rather than using the ${\rm CO}$ or ${\rm H}_2$ abundances determined from the {\sc SGChem} during run-time. We convert these ${\rm CO}$ maps back to synthetic ${\rm H}_2$ maps using a constant ${\rm H}_2$-to-${\rm CO}$ conversion factor $\alpha_{\rm CO} = 4.3~{\rm M}_\odot ({\rm K}~{\rm kms}^{-1}{\rm pc}^{-2})^{-1}$, mimicking the procedures used in observations~\citep{Bolatto13}. This allows a direct comparison of our results in Section~\ref{Sec::molecular-props} to observed molecular cloud populations. Our motivation for this method is that, while {\sc SGChem} produces fully time-dependent chemical abundances, it does not calculate ${\rm CO}$ excitation or line emission, whereas {\sc Despotic} includes a full treatment of the ${\rm CO}$ emission, out of local thermal equilibrium. This allows us to capture the effects of local variations in the ${\rm CO}$ luminosity per unit ${\rm H}_2$ mass, which may be important for comparing to observations. Full details of the post-processing procedure are provided in Appendix~\ref{App::postproc}.

\subsection{Galactic-scale properties of the interstellar medium} \label{Sec::disc-props}
\subsubsection{Disc morphology} \label{Sec::morphology}
The face-on and edge-on gas column densities across all simulation resolutions and feedback prescriptions are displayed in Figure~\ref{Fig::morphology}. In the medium- (centre row) and high-resolution (top row) cases, the addition of momentum from HII regions visibly reduces the sizes of the largest voids in the gas of the interstellar medium, blown by supernova feedback. This corresponds to a qualitative reduction in the amount of outflowing gas from the galactic mid-plane, as seen in the edge-on view, and so to a visible reduction in the gas disc scale-height. The introduction of thermal energy from HII regions without momentum (`HII heat') has no effect on the interstellar medium. In the low-resolution case (bottom row), the difference between the simulations with and without HII region momentum is eradicated. This can likely be attributed to the reduction in supernova clustering with decreasing resolution, as discussed in Section~\ref{Sec::discussion}.

Figure~\ref{Fig::tuning-forks} quantifies the structure of the multi-scale molecular gas distribution in our simulations, relative to the distribution of young stars. This is the result of measuring the gas-to-stellar flux ratio enclosed in apertures centred on ${\rm H}_2$ peaks (top branch) and SFR peaks using `young stars' with ages in the range $0$ to $5$~Myr (bottom branch), and for aperture sizes ranging between $50$~pc and $4000$~pc, following~\cite{Kruijssen2014} and~\cite{Kruijssen18a}. The deviation of the lower branch from the top branch, which sets in at around the gas-disc scale-height~\citep[see also][]{2019Natur.569..519K,Jeffreson21a}, indicates how effectively (on average) molecular gas is removed from around young star clusters in each simulation. If the regions surrounding young stars are effectively cleared of dense gas, then the lower branch drops significantly below the galactic average gas-to-stellar flux ratio at small scales. By contrast, if the young stars remain embedded for long periods of time, then the lower branch remains close to the galactic average value. This is seen in the simulations of~\cite{Fujimoto19}, who find a duration of $23 \pm 1$~Myr, nearly an order of magnitude longer than observed~\citep{2014ApJ...795..156W,2015MNRAS.449.1106H,2018MNRAS.481.1016G,2019MNRAS.483.4707G,2019MNRAS.490.4648H,2019Natur.569..519K,Chevance20,2020arXiv201107772K,2021ApJ...909..121M}. In our simulations, this time-scale ranges from $4.4$~Myr (HII region momentum runs) up to $>5$~Myr (runs without HII region momentum; representing a lower limit, because the duration of co-existence cannot exceed the adopted duration of the young stellar phase, which is $5$~Myr). All of the above numbers are comparable to those obtained for the galaxies with the highest gas surface densities (appropriate for the Agora initial conditions) in the observational sample of~\cite{Chevance20}, who used the same diagnostic to infer time-scales. This provides a qualitative indication that our feedback implementation broadly matches observed feedback-driven dispersal rates of molecular clouds. Indeed, we see that our HII region momentum feedback moves the morphology of the molecular gas and stellar distribution towards that observed in NGC 628. The qualitative result that the top branch is flatter than the bottom branch indicates a cloud lifetime that is longer than the lifetimes of the young stellar groups (here chosen to be 5 Myr). In Section~\ref{Sec::GMC-lifetimes}, we further discuss the influence of our feedback prescription on molecular cloud lifetimes and cloud properties.

\subsubsection{Galactic outflows} \label{Sec::outflows}
The top row of Figure~\ref{Fig::outflows} shows the total galactic star formation rate as a function of the simulation time $t$ at each simulation resolution and for each feedback prescription. At the beginning of the simulation, the disc collapses vertically and a burst of star formation is produced, after which the interstellar medium settles into a state of dynamical equilibrium. In our simulations, equilibrium is achieved after around $200$~Myr. In the medium- and high-resolution cases, the introduction of HII region momentum suppresses the initial starburst at earlier times and so decreases its magnitude. No such effect is seen for the thermal HII regions (`HII heat'), or in any of the low-resolution simulations, mirroring the qualitative results presented in Section~\ref{Sec::morphology}. At $\sim 600$~Myr the star formation rate is consistent with current observed values in the Milky Way~\citep{Murray&Rahman10,Robitaille&Whitney10,Chomiuk&Povich,Licquia&Newman15}. The feedback prescription does not have a perceivable effect on the global star formation rate after the galaxies have equilibriated.

In the centre row of Figure~\ref{Fig::outflows}, we show the rate of gas outflow from each galaxy. The outflow rates are calculated as the total momentum of the gas moving away from the disc, summed over two planar slabs of thickness $500~{\rm pc}$, located at $\pm 5$~kpc above and below the galactic disc. This is the same definition used in~\cite{2014MNRAS.442.3013K,2020arXiv200403608K}. In the medium- and high-resolution simulations, the outflow rate is decreased by around an order of magnitude upon the introduction of HII region momentum feedback. This is again consistent with a reduced level of supernova clustering, which decreases the effectiveness of supernova feedback in driving outflows~\citep{2020arXiv200911309S,2020arXiv200403608K}. The mass-loading $\eta$ of the stellar feedback in our model (bottom row of Figure~\ref{Fig::outflows}) divides the outflow rate by the star formation rate. We note that there is a clear resolution-dependence of the feedback-induced outflow rates and mass-loadings for all feedback prescriptions, likely due to the increased clustering of supernovae at higher resolutions. This is discussed further in Section~\ref{Sec::SNe-fb}.

\subsubsection{Resolved disc stability} \label{Sec::vss}
The presence of momentum feedback from HII regions makes a significant difference to the velocity dispersion $\sigma_{\rm g}$ and gravitational stability $Q_{\rm g}$ of the cold gas ($T \leq 10^4$~K) in our high- and medium-resolution simulations (left and centre columns in Figure~\ref{Fig::vss}, respectively), as well as to the scale-height $h_{\rm g}$ of the total gas distribution. We calculate the line-of-sight turbulent velocity dispersion as
\begin{equation}
\sigma_{\rm los, g}^2 = \frac{\langle |\bm{v}_i - \langle \bm{v}_i \rangle|^2 \rangle}{3},
\end{equation}
where $\{\bm{v}_i\}$ are the velocity vectors of the gas cells in each radial bin, and angled brackets denote mass-weighted averages over these cells. The~\cite{Toomre64} $Q$ parameter of the cold gas is then defined as
\begin{equation}
Q_{\rm g} = \frac{\kappa \sigma_{\rm g}}{\pi G \Sigma_{\rm g}},
\end{equation}
with $\kappa$ the epicyclic frequency of the galactic rotation curve and $\sigma_{\rm g} = \sqrt{c_s^2 + \sigma_{\rm los, g}^2}$ for gas sound speed $c_s$. In the top row of Figure~\ref{Fig::vss}, we quantitatively show the result for the disc scale-height that was demonstrated qualitatively in Figure~\ref{Fig::morphology}: the reduction in the violence of feedback-induced outflows perpendicular to the galactic mid-plane leads to a smaller disc scale-height when momentum from HII regions is incorporated. In the second row, we demonstrate that for galactocentric radii $R < 8$~kpc in the high-resolution simulation, the amount of cold gas is increased by up to 50~per~cent when HII region momentum is included (solid lines) and that the amount of molecular gas is almost doubled (dashed lines). This is due to two effects: (1) the overall mass of the interstellar medium is larger in the simulations with HII region momentum, due to the suppression of the initial `starburst' (see Section~\ref{Sec::outflows}), and (2) the fraction of the interstellar medium in the cold and molecular phases is increased (see Section~\ref{Sec::phases}). We also find that the cold gas has a lower velocity dispersion by $\sim 5~{\rm kms}^{-1}$ at all galactocentric radii. Accordingly, the Toomre $Q$ factor ($Q_{\rm g}$, bottom row of panels) is suppressed by a factor of $\sim 2$ out to $R \sim 8~{\rm kpc}$. The HII region momentum causes the interstellar medium to become clumpier and less gravitationally-stable, leading to the formation of more molecular clouds, as will be discussed in Section~\ref{Sec::molecular-props}. This is again consistent with the idea that the HII region feedback reduces the momentum injected by supernova feedback, likely by reducing its clustering. In the low-resolution case, none of the observables associated with galactic disc stability are altered by the addition of HII region momentum, consistent with the results presented in Sections~\ref{Sec::morphology} and~\ref{Sec::outflows}.

%========================= PHASES
\begin{figure}
  \label{Fig::phases}
    \includegraphics[width=\linewidth]{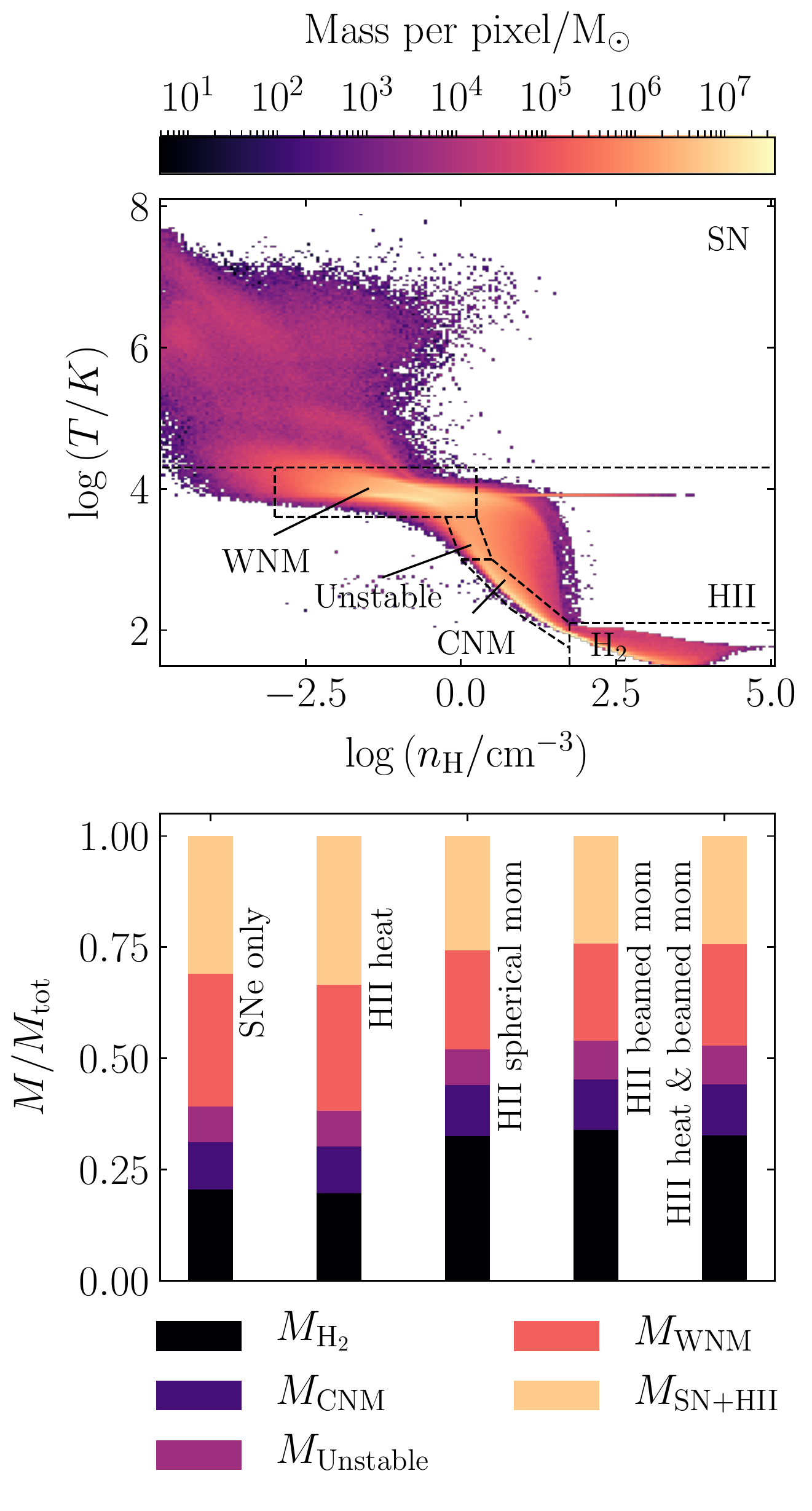}
    \caption{{\it Upper panel:} Density-temperature phase diagram for the HI-resolution simulation with both thermal and beamed momentum from HII regions (HII thermal + beamed mom.), at $t=600$~Myr. Dashed lines delineate the regions of phase space corresponding to the warm neutral medium (WNM), the thermally-unstable phase (Unstable), the cold neutral median (CNM), gas heated by HII regions (HII), and gas heated by supernovae (SN). {\it Lower panel:} Partitioning of the gas mass in each HI-resolution simulation into five ISM phases from warmest to coolest, as a fraction of the total gas mass in the simulation: hot gas that has received thermal energy from stellar feedback ($M_{\rm SN+HII}$), the warm neutral medium ($M_{\rm WNM}$), the unstable phase ($M_{\rm unstable}$), the cold neutral medium ($M_{\rm CNM}$), and the star-forming gas in the molecular phase ($M_{\rm H_2}$, as computed using {\sc Despotic}, see Appendix~\protect\ref{App::postproc}).}
\end{figure}
%=========================

%========================= KS RLNS
\begin{figure*}
  \label{Fig::KS-rln}
    \includegraphics[width=\linewidth]{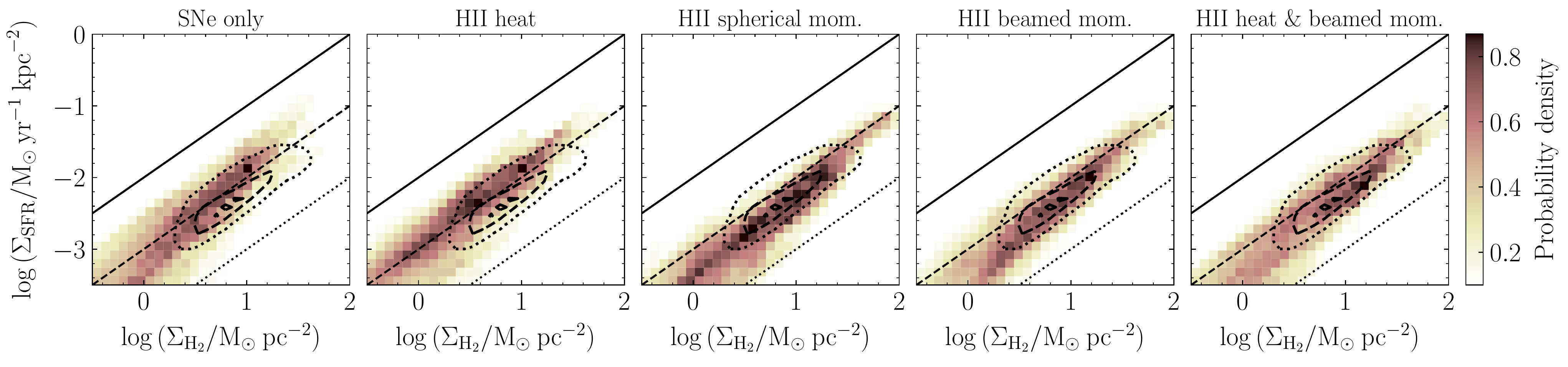}
    \caption{Pixel density as a function of $\Sigma_{\rm H_2}$ and $\Sigma_{\rm SFR}$ each disc, at a spatial resolution of $750$~pc, corresponding to the resolved molecular star-formation relation of~\protect\cite{Kennicutt98}. Gas depletion times of $10^8$, $10^9$ and $10^{10}$~Myr are given by the black solid, dashed and dotted lines respectively. The contours encircle 90~per~cent (dotted), 50~per~cent (dashed) and 10~per~cent (solid) of the observational data for nearby galaxies from~\protect\cite{Bigiel08}. All maps are computed at $t=600$~Myr.}
\end{figure*}
%=========================

%========================= CLOUD LIFETIMES
\begin{figure}
  \label{Fig::cloud-lifetimes}
    \includegraphics[width=\linewidth]{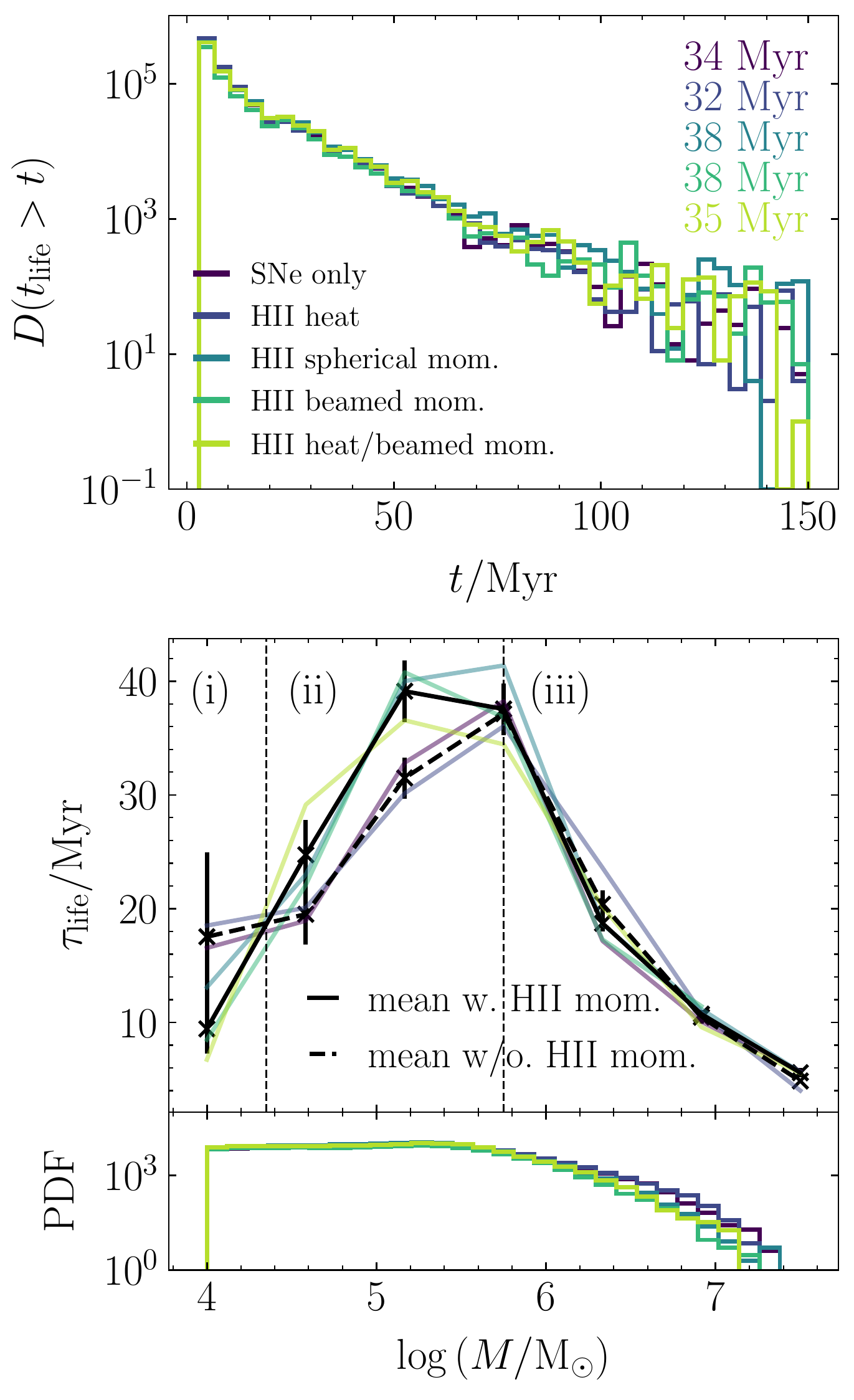}
    \caption{\textit{Top:} Cumulative distribution of trajectory lifetimes $t_{\rm life}$ in each of the high-resolution simulations. The characteristic cloud lifetimes, obtained from the exponential distributions by fitting a function $\exp{(-t/\tau_{\rm life})}$ according to Equation~(\ref{Eqn::lifetime-dstbn}), are annotated according to the legend colours. \textit{Bottom:} Characteristic cloud lifetime for each simulation (transparent solid lines) as a function of the cloud mass, with the cloud mass PDFs below. The mean values with and without HII region momentum feedback are given by the solid and dashed black lines, respectively. The error-bars correspond to the errors associated with the exponential fits to the distributions $D(t_{\rm life}>t)$ in each mass bin. Three regimes are annotated: $(i)$ for clouds destroyed preferentially by HII region feedback, $(ii)$ for clouds destroyed preferentially by supernovae, and $(iii)$ for clouds dominated by interactions.}
\end{figure}
%=========================

%========================= NODE CONNECTIVITY
\begin{figure}
  \label{Fig::node-connectivity}
    \includegraphics[width=\linewidth]{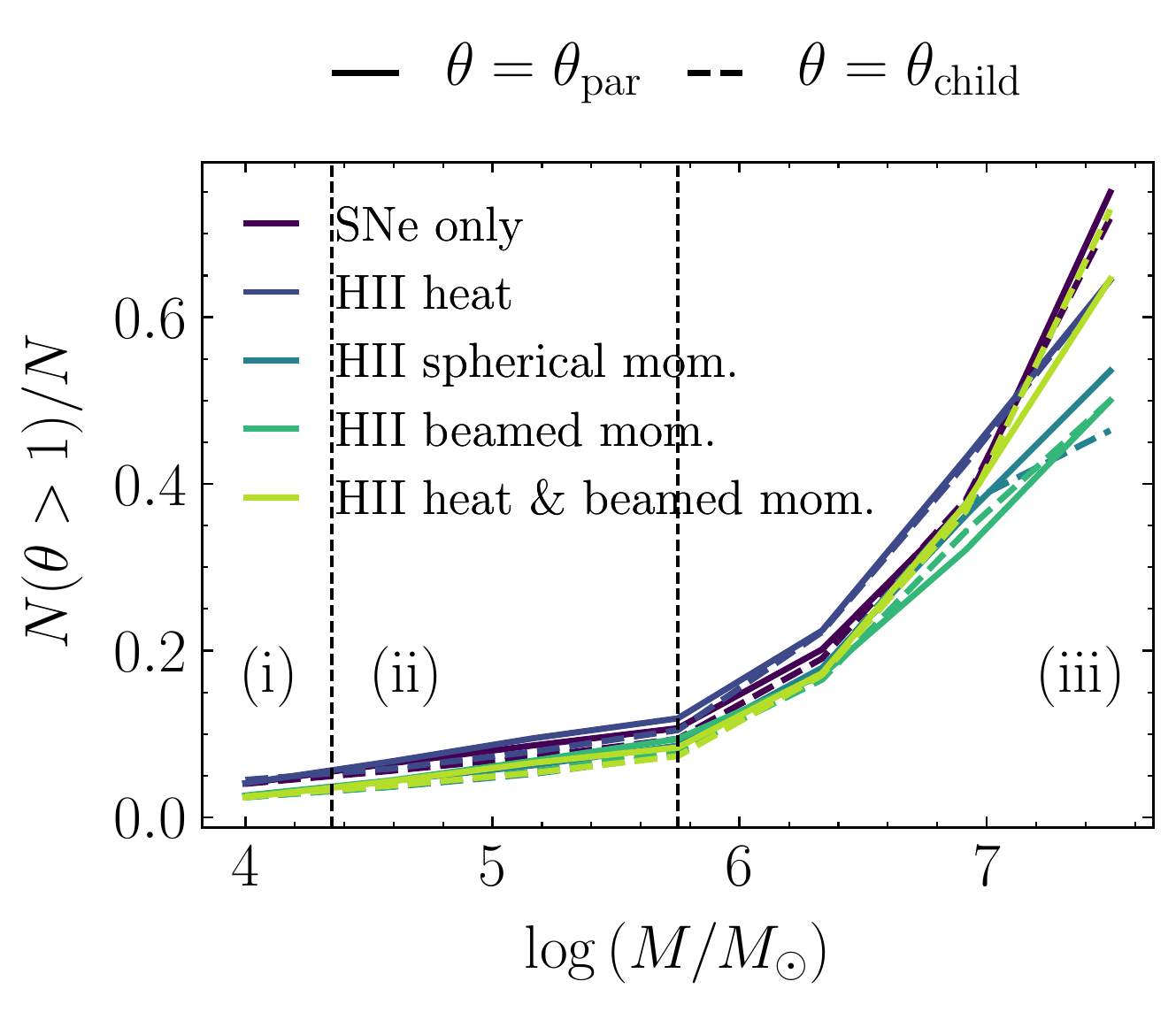}
    \caption{Fraction of nodes in the cloud evolution network (molecular clouds observed at an instant in time) that split into two or more children ($\theta_{\rm child}>1$) or are the result of a merger between two or more parents ($\theta_{\rm par}>1$), as a function of cloud mass. We see that the connectivity of the network increases exponentially from $\sim 10$~per~cent of multiply-connected nodes for cloud masses $M \sim 5\times 10^5 M_\odot$ up to $\sim 70$~per~cent at $M \sim 3\times 10^7 M_\odot$. This is the same mass range over which cloud lifetimes decrease with mass in the lower panel of Figure~\protect\ref{Fig::cloud-lifetimes}, and cease to depend on the feedback prescription used. The three mass regimes are annotated as in Figure~\protect\ref{Fig::cloud-lifetimes}.}
\end{figure}
%=========================

\subsubsection{ISM phase structure} \label{Sec::phases}
In the top panel of Figure~\ref{Fig::phases} we display the mass-weighted distribution of gas temperature as a function of the gas volume density (the phase diagram) for the high-resolution simulation including both thermal and beamed HII region momentum (`HII heat \& beamed mom'). The gas cells cluster around a state of thermal equilibrium in which the rate of cooling (dominated in our simulations by line emission from ${\rm C}^+$, ${\rm O}$ and ${\rm Si}^+$) balances the rate of heating due to photoelectric emission from PAHs and dust grains. The thin horizontal line of particles at high volume densities and $T \sim 7000$~K contains the particles that are heated by the thermal feedback from HII regions. The dashed black lines delineate the partitioning of the interstellar medium into the feedback-heated phases (SN and HII) the warm neutral medium (WNM), the unstable phase, the cold neutral medium (CNM) and the set of gas cells that are predominantly molecular (${\rm H}_2$). We have chosen the partitioning of the WNM and CNM gas by eye, according to the major regions of gas accumulation along the thermal equilibrium curve in the phase diagram. The region bridging the WNM and CNM is then classified as `unstable' following~\cite{Goldbaum16}, and material that is lifted above the equilibrium curve is attributed to feedback-related heating. In the lower panel of Figure~\ref{Fig::phases} we show the fraction of the total gas mass in each of these phases for the five high-resolution simulations. The mass of molecular hydrogen we use is that which would be inferred by an observer from the CO luminosity, as computed by {\sc Despotic} (see Appendix~\ref{App::postproc}). The addition of thermal feedback from HII regions does nothing to the phase structure of the interstellar medium, relative to the case of supernovae only. By contrast, explicit injection of momentum from HII regions leads to almost double the mass of molecular gas and $\sim 50$~per~cent more cold gas overall ($T \leq 10^4$~K). The masses of warm and hot, feedback-heated gas are correspondingly reduced. We also note that the overall gas mass remaining in the galaxy at $t=600$~Myr is larger by around $0.8 \times 10^9~{\rm M}_\odot$. This is because the initial `bursts' of star formation, as the galaxy settles into equilibrium, are smaller in the case of effective pre-supernova feedback, as discussed in Section~\ref{Sec::outflows}.

\subsubsection{Star formation in molecular gas} \label{Sec::molecular-SF}
Although the global star formation rate in our simulations appears insensitive to the feedback prescription applied (top row of Figure~\ref{Fig::outflows}), a slightly greater level of variation is revealed when we look explicitly at the star formation rate surface density $\Sigma_{\rm SFR}$ as a function of the molecular gas surface density $\Sigma_{\rm H_2}$ (the molecular Kennicutt-Schmidt relation) in Figure~\ref{Fig::KS-rln}. We find that with the addition of momentum feedback from HII regions, the gradient of the slope in the $\Sigma_{\rm H_2}$-$\Sigma_{\rm SFR}$ plane is flattened slightly and the molecular gas surface densities are increased by a factor of around two. This means that they fall closer to the observed values delineated by the closed black contours. However this fact should not be over-interpreted,  given that the size of the shift in surface density is smaller than the uncertainty in the ${\rm H}_2$-to-CO conversion factor used to compute the molecular gas abundances (see Appendix~\ref{App::postproc}). Again, the addition of thermal HII region feedback on its own has no effect.

\subsection{Properties of molecular clouds} \label{Sec::molecular-props}
In this section, we analyse the molecular clouds identified at the native spatial resolution ($6$~pc) of the column-density projections for our high-resolution simulations. These clouds span a size range from $18$~pc up to $200$~pc and a mass range from $100~{\rm M}_\odot$ up to $10^8~{\rm M}_\odot$. We identify clouds by taking a threshold of $\log{(\Sigma_{\rm H_2}/{\rm M}_\odot~{\rm pc}^{-2})} = -3.5$ on the molecular gas column density, as calculated using {\sc Despotic} (see the beginning of Section~\ref{Sec::results} and Appendix~\ref{App::postproc}). The clouds themselves are identified using the {\sc Astrodendro} package for Python. This procedure is described in detail in Appendix~\ref{App::postproc}, and is discussed at length in Section 2.9 of~\cite{2020MNRAS.498..385J}, where we also show that the molecular clouds identified by this method have properties in agreement with observations of clouds in Milky Way-like galaxies, including their masses, sizes, velocity dispersions, surface densities, pressures and star formation rate surface densities. Similarly to~\cite{2020MNRAS.498..385J}, we discard clouds spanning fewer than 9 pixels ($324~{\rm pc}^2$), or containing fewer than 20 Voronoi gas cells.

Once the molecular clouds in our simulations have been identified at every simulation time-step, we follow their evolution as a function of time according to the procedure described in Section 3.2 of~\cite{Jeffreson21a}. Briefly, we take the two-dimensional pixel masks associated with the sets of molecular clouds in two consecutive snapshots at times $t=t_1$ and $t=t_2$. We project the mask positions of the clouds at $t=t_1$ using the positions and velocities of the gas cells that they span, such that $(x_1, y_1) \rightarrow (x_1 + v_x \Delta t, y_1 + v_y \Delta t)$. We then compare the projected masks to the true pixel masks of the clouds at $t=t_2$. If the projected and true pixel maps overlap by one pixel, then the clouds are indistinguishable at the spatial resolution ($6$~pc) and temporal resolution ($1$~Myr) of the snapshots, and so each cloud at $t=t_1$ is assigned as a parent of the children at $t=t_2$. A given cloud can spawn multiple children (\textit{cloud splits}) or have multiple parents (\textit{cloud mergers}). We store the network of parents and children using the {\sc NetworkX} package for python~\citep{NetworkX}, and `prune away' unphysical nodes produced by regions of faint CO background emission in our astrochemical post-processing, which do not contain sufficient quantities of CO-luminous gas. We find that these nodes can be removed by taking a cut of $\sigma \sim 0.03~{\rm kms}^{-1}$ on the cloud velocity dispersion, as described in~\cite{Jeffreson21a}.\footnote{The mass cut applied in~\cite{Jeffreson21a} is not required here, as we discard clouds containing fewer than 20 Voronoi cells.}

\subsubsection{Molecular cloud lifetimes} \label{Sec::GMC-lifetimes}
Using the \textit{cloud evolution network} described above, we calculate the lifetime $t_{\rm life}$ of each distinct molecular cloud identified at a given time in our simulations, by performing a Monte Carlo (MC) walk through the network. At the beginning of each MC iteration, walkers are initialised at every \textit{formation node} in the network (nodes corresponding to a net increase in cloud number). The walkers step along time-directed edges of length $1$~Myr between consecutive nodes, until an \textit{interaction node} is reached. An interaction may be a merger, a split, or a transient meeting. A random number from a uniform distribution is used to choose between the possible subsequent trajectories for each walker, including the possibility of cloud destruction, if it exists at that node. If the cloud is destroyed, the final lifetime $t_{\rm life}$ is returned. This algorithm satisfies the requirements of:
\begin{enumerate}
	\item \textit{Cloud uniqueness:} Edges between nodes in the network represent time-steps in the evolution of a single cloud, so must not be double-counted.
	\item \textit{Cloud number conservation:} The number of cloud lifetimes retrieved from the network must be equal to the number of cloud formation events and cloud destruction events, as each cloud can be formed and destroyed just once.
\end{enumerate}
Seventy MC iterations are performed to reach convergence of the characteristic molecular cloud lifetime $\tau_{\rm life}$ for the cloud population of the entire galaxy.

In the top panel of Figure~\ref{Fig::cloud-lifetimes}, we show the cumulative distributions $D(t_{\rm life} > t)$ of lifetimes $t_{\rm life}$ for the molecular clouds in our high-resolution simulations. These distributions have an exponential form, as expected if the formation and destruction of clouds has reached a steady state. The simulations with HII region momentum feedback do not appear significantly different to those without. We have annotated the \textit{characteristic cloud lifetime} $\tau_{\rm life}$ for each simulation by fitting an exponential profile to each distribution, and assuming the steady-state proportionality
\begin{equation} \label{Eqn::lifetime-dstbn}
\ln{D(t_{\rm life})} \propto -\frac{t}{\tau_{\rm life}},
\end{equation}
as in~\cite{Jeffreson21a}. We find only a marginal increase of $4$~Myr in the overall value of $\tau_{\rm life}$ upon the introduction of HII region momentum feedback (an average of $37 \pm 2$~Myr with HII region momentum vs. $33 \pm 2$~Myr without). However, in the lower panel of Figure~\ref{Fig::cloud-lifetimes}, we see that the the influence of the feedback prescription is dependent on the cloud mass. Its influence can be divided into three regimes as follows:
\begin{enumerate}
	\item $M/{\rm M}_\odot \la 5.6 \times 10^4$: HII region momentum depresses the cloud lifetime by $\sim 10$~Myr.
	\item $5.6 \times 10^4 \la M/{\rm M}_\odot \la 5 \times 10^5$: HII region momentum increases the cloud lifetime by $\sim 7$~Myr.
	\item $5 \times 10^5 \la M/{\rm M}_\odot$: HII region momentum has no effect on the cloud lifetime.
\end{enumerate}
This result is consistent with the following scenario: the least massive molecular clouds in $(i)$ are less likely to contain the massive stellar clusters required for the fastest and most efficient injection of supernova energy. This results in an uptick of the characteristic cloud lifetime for the simulations without HII region momentum feedback (blue and purple lines in Figure~\ref{Fig::cloud-lifetimes}) at small masses. However, the least-massive clouds are also the easiest to disperse, and so the relatively-small amount of momentum injected by HII regions can truncate the cloud lifetime in the absence of efficient supernova feedback. At larger cloud masses $(ii)$, the HII region momentum is too puny to cause disruption, so its main influence is to reduce supernova clustering and thus decrease the efficacy of the supernova feedback, consistent with its effect on the large-scale properties of the interstellar medium, presented in Section~\ref{Sec::disc-props}. This increases the characteristic cloud lifetime. Finally, the most massive molecular clouds in $(iii)$ are often unresolved cloud complexes, and undergo increasingly more mergers and splits as the cloud mass is increased from $10^{5.8}$ through $10^{7.5}~{\rm M}_\odot$, as shown in Figure~\ref{Fig::node-connectivity}. Across this mass regime, the fraction of multiply-connected nodes increases from 10~per~cent up to 70~per~cent, elevating the number of short MC trajectories containing high-mass nodes. The trajectory lifetimes returned by the MC walk are therefore likely to be determined by the level of graph connectedness, rather than by the feedback-induced destruction of the molecular clouds. This also explains the drop in the cloud lifetime for the most massive clouds. In order to determine the effects of stellar feedback on these high-mass clouds, we will need to develop the algorithm put forward in~\cite{Jeffreson21a}, to distinguish between cloud mergers of varying mass ratio. Overall, the cloud lifetimes across masses span the range from $10$-$40$~Myr, similar to observations~\citep{Engargiola03,Blitz2007,Kawamura09,Murray11,Meidt15,Corbelli17,Chevance20}.

Finally, we note that the number of molecular clouds generated per unit mass of the interstellar medium in our simulations is increased by the presence of HII region momentum. In the `SNe only' and `HII heat' simulations, the average number of clouds identified is $3.7$ per $10^7~{\rm M}_\odot$; this increases to $6.2$ per $10^7~{\rm M}_\odot$ for the `HII spherical mom.', `HII beamed mom.' and `HII heat \& beamed mom.' simulations. Combined with the reduced number of high-mass clouds, this result indicates that the molecular interstellar medium is slightly more fragmented in the case of the HII region feedback.

%========================= CLOUD VELDISPS AND SURFDENSES
\begin{figure*}
  \label{Fig::cloud-veldisp-surfdens}
    \includegraphics[width=\linewidth]{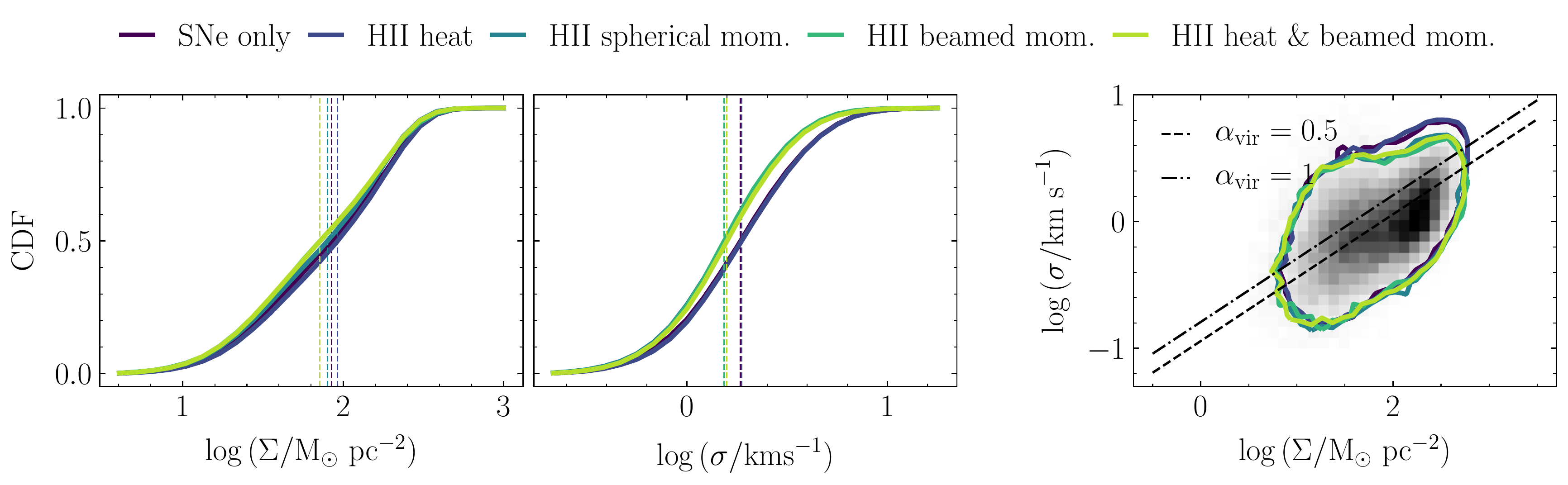}
    \caption{Distributions of molecular cloud surface densities $\Sigma$ and velocity dispersions $\sigma$ in each of the high-resolution simulations. Each value is a median over a trajectory in the cloud evolution network. \textit{Left/Centre:} Cumulative distribution of the surface density/velocity dispersion, with medians indicated by the vertical dashed lines. \textit{Right:} Scaling relation of the surface density with the velocity dispersion. The contours enclose 90~per~cent of the identified molecular clouds. The grey-shaded histogram contains the full cloud distribution for the `HII heat \& beamed mom.' simulation. Virial parameters of $\alpha_{\rm vir}=0.5$ and $1$ for spherical clouds of size $6$~pc are given by the dashed and dot-dashed lines, respectively. We see that the introduction of HII region momentum predominantly (but hardly) affects the cloud velocity dispersion.}
\end{figure*}
%=========================

\subsubsection{Cloud velocity dispersions and surface densities} \label{Sec::GMC-veldisp-surfdens}
We now turn to the physical properties of the molecular clouds in our simulations: first to the scaling relation between the cloud surface density $\Sigma$ and velocity dispersion $\sigma$. Each value corresponds to an average (median) taken over the cloud lifetime $t_{\rm life}$ (i.e.~along a unique trajectory in the cloud evolution network). The right-hand panel of Figure~\ref{Fig::cloud-veldisp-surfdens} shows the scaling relation itself, for which the clouds fall along a line of constant virial parameter, as observed in nearby Milky Way-like galaxies~\citep[e.g.][]{Sun18}. Lines representing virial parameters of $1$ and $2$ for spherical clouds at a fixed size of $6$~pc (our native resolution) are given by the dashed and dot-dashed lines, respectively. The coloured contours enclose 90~per~cent of the clouds for each high-resolution simulation, while the grey-shaded histogram displays the entire cloud population for the `HII heat \& beamed mom.' simulation. In the left and central panels of Figure~\ref{Fig::cloud-veldisp-surfdens} we show the cumulative distributions of the cloud surface density and velocity dispersion separately. We see that the introduction of HII region momentum makes little difference to the distribution of surface densities, and reduces the median cloud velocity dispersion by only $0.5~{\rm kms}^{-1}$. This reduction is consistent with the drop in the bulk velocity dispersion of the cold gas in our simulations, presented in Figure~\ref{Fig::vss}.

\subsubsection{Cloud masses and star formation rates} \label{Sec::GMC-mass-SFR}
\begin{figure*}
  \label{Fig::cloud-mass-SFR}
    \includegraphics[width=\linewidth]{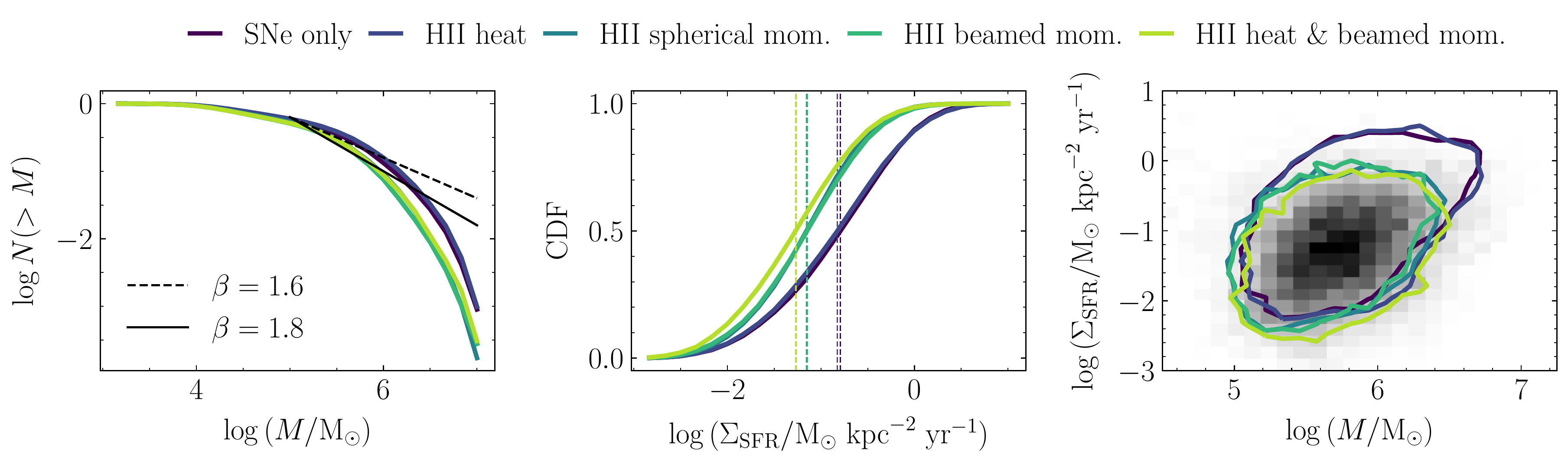}
    \caption{Distributions of molecular cloud masses $M$ and star formation rates per unit area of the galactic mid-plane $\Sigma_{\rm SFR}$ in each of the high-resolution simulations. \textit{Left:} Normalised number of clouds $N(>M)$ with mass larger than or equal to $M$. The solid black and dashed black lines denote the range of power-law slopes for the observed cloud mass distribution in the Milky Way, given by $\dd N/\dd M \propto M^{-\beta}$ with $\beta \in [1.6, 1.8]$. \textit{Centre:} Cumulative distribution of the star formation rate surface density, with medians indicated by the vertical dashed lines. \textit{Right:} Scaling relation of the star formation rate surface density with the cloud mass. The contours enclose 90~per~cent of the identified molecular clouds. The grey-shaded histogram contains the full cloud distribution for the `HII heat \& beamed mom.' simulation. We see that the introduction of HII region momentum reduces the star formation rates of all clouds, and slightly decreases the number of massive clouds.}
\end{figure*}

The influence of the stellar feedback prescription on the masses and star formation rate surface densities of our molecular clouds is shown in Figure~\ref{Fig::cloud-mass-SFR}. In the right-hand panel, the number of clouds $N(>M)$ with mass greater than $M$ is compared to the power-law form $\dd N/\dd M \propto M^{-\beta}$ observed for clouds in the Milky Way over the mass range of $\log{M} \in [4.8, 6.5]$, with $\beta \in [1.6, 1.8]$~\citep{Solomon87,Williams&McKee97,Heyer+09,Roman-Duval+10,Miville-Deschenes17,Colombo+19}. When we fit corresponding powerlaws to the PDF of each mass spectrum (via simple linear regression in the mass range $\log{(M/M_\odot)} \in [4.8, 6.5]$), we find a slope of $\beta = 1.75 \pm 0.09$ for the `SNe only' simulation and a slope of $\beta = 1.80 \pm 0.12$ for the `HII heat \& beamed mom.' simulation. That is, the number of the most massive clouds is reduced slightly by the presence of HII region feedback.

We note that this result (a steeper mass function with HII region momentum) is the opposite of that expected if the characteristic rates $\xi_{\rm form}$ and $1/\tau_{\rm life}$ for cloud formation and destruction in each galaxy are independent of the cloud mass, as assumed in~\cite{2017ApJ...836..175K}. These authors use an analytic rate equation for the number of clouds $N$, explicitly accounting for the process of cloud coagulation, to derive a mass function slope of $\dd N/\dd M \propto -(\xi_{\rm form} \tau_{\rm life})^{-1}$. In the steady-state approximation of Equation (\ref{Eqn::lifetime-dstbn}), the number of clouds present in the galaxy at a given time approaches $N \rightarrow \tau_{\rm life} \xi_{\rm form}$, so the predicted slope goes as $\dd N/\dd M \propto -1/N$. We find $N$ to be higher in the simulations with HII region momentum, but $\dd N/\dd M$ to be steeper, in contradiction with this work. We attribute this to the fact that $\tau_{\rm life}$ is manifestly dependent on the cloud mass (see Figure~\ref{Fig::cloud-lifetimes}) and that the mass-dependence of $\xi_{\rm form}$ is not studied here, but likely non-negligible.

In the central panel of Figure~\ref{Fig::cloud-mass-SFR}, we show the star formation rate $\Sigma_{\rm SFR}$ per unit area of the molecular clouds in each simulation. The introduction of HII region momentum causes a three-fold drop in the value of $\Sigma_{\rm SFR}$. In the right-hand panel, we show that this drop in the star formation rate occurs across the whole range of cloud masses. This result agrees broadly with the results from high-resolution simulations of resolved HII regions~\citep[e.g.][]{2016ApJ...829..130R,2018MNRAS.478.4799H,2018MNRAS.475.3511G,2019MNRAS.489.1880H,2020MNRAS.497.3830F,2020MNRAS.499..668G,2020MNRAS.492..915G,2020arXiv201107772K}, which show that HII region feedback can efficiently suppress the overall star formation efficiency within individual molecular clouds.

%========================= CLOUD MASSES AND SFRS
%=========================

\subsection{Beamed vs.~spherical HII region momentum} \label{Sec::beamed-vs-spherical}
Aside from the finding that thermal feedback from marginally-resolved HII regions is ineffective in transferring energy to the surrounding interstellar medium, a recurring theme in the preceding sub-sections is that there is no discernible difference between our simulations with spherical HII region feedback and beamed HII region feedback. The morphology, phase structure and stability of the interstellar medium are identical in these cases, and the properties of the molecular clouds are unaffected. This might be surprising, considering the qualitative difference in the appearance of HII regions in the blistered and spherical cases (see Figure~\ref{Fig::blister-projections}) and the difference in their ionisation-front and Str\"{o}mgren radii. We find that it is only the quantity of momentum injected in our simulations that matters (this is roughly equivalent in the spherical and beamed cases), and not the direction in which it is injected. However, we might expect that if the direction of momentum injection were not chosen randomly for each FoF group and star particle, but rather preserved over the evolution of each molecular cloud, the blistered HII region feedback might be more effective in removing the gas from around star particles.

\section{Discussion} \label{Sec::discussion}
We have shown in Section~\ref{Sec::results} that the injection of momentum from HII regions, according to a novel numerical model based on the analytic framework of KM09, reduces the mass-loading of outflows perpendicular to the mid-plane of isolated disc galaxies, and increases the fraction of cold gas within these discs, while decreasing its velocity dispersion and scale-height. The resolved molecular clouds formed from this cold gas reservoir suffer alterations in their lifetimes, masses, star formation rates and velocity dispersions. We find that these results apply across a mass resolution range of $10^3$-$10^4~{\rm M}_\odot$ in the moving-mesh code {\sc Arepo.}

It is important to note that all of the above results depend not just on our modelling of HII regions, but on a number of other assumptions made during the construction of our stellar population and its feedback, including its supernova feedback. In Section~\ref{Sec::caveats}, we outline the key caveats of our model, their possible effects, and how these could in the future be disentangled from the relative roles of HII region and supernova feedback in isolated galaxy simulations. In Section~\ref{Sec::literature-sims} we compare our results to studies of HII region feedback in the literature.

\subsection{Caveats of our model} \label{Sec::caveats}
\subsubsection{Photon trapping and escape} \label{Sec::comp-highres}
Within the model for HII region feedback, we have used a value of
\begin{equation}
f_{\rm trap} = 1 + f_{\rm trap, w} + f_{\rm trap, IR} + f_{{\rm trap, Ly}\alpha} \sim 8
\end{equation}
to account for the enhancement of pressure inside the ionisation front, due to the trapping of energy from stellar winds ($f_{\rm trap, w}$), infrared photons ($f_{\rm trap, IR}$) and Lyman-$\alpha$ ($f_{{\rm trap, Ly}\alpha}$) photons. The value of $f_{\rm trap}$ is constrained by~\cite{2021ApJ...908...68O} using infrared observations to infer the pressures inside young HII regions in the Milky Way. By using $f_{\rm trap} \sim 8$, we therefore implicitly assume that $f_{{\rm trap, Ly}\alpha} \approx 0$ and $f_{\rm trap, w} \approx 0$, because an estimation of the effects of Lyman-$\alpha$ photons and winds would require observations of the dust and diffuse gas surrounding the sources, in the optical and the X-ray, respectively. Our model therefore does not account for the absorption of Lyman-$\alpha$ photons, or for the trapping of stellar winds. In addition, the interaction of stellar winds with radiation pressure is a complex problem: numerous high-resolution, radiative-transfer numerical studies of HII regions inside individual clouds~\citep[e.g.][]{2017MNRAS.467.1067D,2017ApJ...850..112R,2018ApJ...859...68K,2019ApJ...883..102K} have shown that stellar winds (along with inhomogeneities in the gas surrounding HII regions) can lead to the escape of radiation through holes in the shell bounding the ionised gas, and so to a reduction in the overall radiation pressure by factors of $\sim 5$-$10$~\citep{2019ApJ...883..102K}. The HII regions in our model are assumed perfectly spherical or hemi-spherical, and we have not accounted for stellar winds. It is therefore possible that we have under-estimated the strength of radiation pressure by ignoring Lyman-$\alpha$ photon and wind trapping, or have over-estimated it by ignoring photon escape. However, this is unlikely to have a large effect on the total amount of momentum injected by our HII regions, because for the ionising luminosities between $S_{49} \sim 1$ and $100$ spanned by the FoF groups in our simulations, the momentum contribution made by the gas pressure is around ten times that made by the radiation pressure, according to our Equation~(\ref{Eqn::momentumeqn_final}).

\subsubsection{Resolution-dependence of supernova feedback} \label{Sec::SNe-fb}
As noted throughout Section~\ref{Sec::results}, the differences between the simulations with and without HII region momentum do not persist down to resolutions of $10^5~{\rm M}_\odot$ per gas cell. In addition, when supernova feedback is used on its own, the outflow rates and their mass-loadings, as well as the gravitational stabilities of the gas in the galactic discs, are substantially different for the low-, medium- and high-resolution simulations. This may be due to a decrease in the effectiveness of supernova clustering at resolutions of $<10^5~{\rm M_\odot}$ (i.e. the resolution is too low for clustering to be resolved). As discussed by~\cite{2020arXiv200911309S}, early feedback from HII regions and stellar winds affects the interstellar medium by reducing the degree of supernova clustering and so the violence of the resulting explosions, decreasing the sizes of galactic outflows and the mid-plane gas velocity dispersion. Therefore, if clustering is not resolved, the effect of our HII region feedback on the large-scale properties of the interstellar medium may be spuriously-weakened at low resolutions.~\cite{2020arXiv201010533S} also discuss the non-convergence of various stellar feedback prescriptions due to an under-sampling the IMF at high mass resolutions. However, this is not a problem in our simulations, due to the use of the Poisson sampling procedure from~\cite{Krumholz15}. By this procedure, the number of stars assigned to a given star particle/cluster depends on the star particle mass, but the form of the resulting distribution of stellar masses is not affected.

\subsection{Comparison to the literature} \label{Sec::literature-sims}
\subsubsection{High-resolution simulations of molecular clouds}
The molecular cloud sample in our simulations has yielded two key results: (1) the lifetimes of the least-massive clouds are truncated by HII region feedback (while those of intermediate-mass clouds are extended), and (2) HII region feedback suppresses the star formation rate within molecular clouds by a factor of three. These findings can be qualitatively compared to high-resolution simulations of resolved HII regions in individual molecular clouds. In particular,~\cite{2012MNRAS.424..377D,2013MNRAS.430..234D,2017ApJ...851...93K} find that only the least-massive molecular clouds are prone to dispersal by HII region feedback, and that this dispersal occurs on time-scales of $<10$~Myr, as we have found in Section~\ref{Sec::GMC-lifetimes}. Larger clouds can only be disrupted by supernovae. Across molecular clouds of all masses,~\cite{2016ApJ...829..130R,2018MNRAS.478.4799H,2018MNRAS.475.3511G,2019MNRAS.489.1880H,2020MNRAS.497.3830F,2020MNRAS.499..668G,2020MNRAS.492..915G,2020arXiv201107772K} show that the star formation efficiency per free-fall time is suppressed by the presence of HII region feedback, as we have discussed in Section~\ref{Sec::GMC-mass-SFR}. Although it will be important to check the convergence of our sub-grid model with high-resolution simulations such as these, it is encouraging to note that the main results for our molecular cloud sample echo existing results in single-cloud studies.

\subsubsection{Isolated disc simulations}
We may also compare our results with other implementations of radiation/thermal pressure from HII regions in isolated disc galaxies at similar mass resolutions.~\cite{2020arXiv200911309S} investigate the role of pre-supernova feedback in suppressing supernova clustering in dwarf galaxies in {\sc Arepo}, reaching mass resolutions of $20~{\rm M}_\odot$. The Str\"{o}mgren radii of the HII regions in their simulations are well-resolved, allowing for the explicit ionisation and heating of gas cells to be converted to momentum. Using this prescription, the authors find that supernova clustering is decreased by the presence of HII region feedback. This leads to a significant suppression of outflows and their mass-loadings, as well as a reduction in the sizes of supernova-blown voids within the interstellar medium, in agreement with our results.~\cite{Fujimoto19} investigate molecular clouds in an isolated disc galaxy at a comparable resolution to ours, but with only thermal HII region feedback~\citep[see also][]{Goldbaum16}. These authors find that both the pre-supernova and supernova feedback in their simulations are inefficient at disrupting the parent molecular clouds around young stars, resulting in a flat scale-dependence of the gas-to-stellar flux ratio when apertures are centred on young stellar peaks (by comparison to the diverging branch we find in our Figure~\ref{Fig::tuning-forks}). This leads to a much longer duration of the embedded phase of star formation, as derived via the method of~\cite{Kruijssen18a}: $23$~Myr in~\cite{Fujimoto19} vs. $4.4$~Myr in our simulations.

At mass resolutions of $10^3$-$10^4$ solar masses per gas cell, ~\cite{2011MNRAS.417..950H,2013MNRAS.434.3142A,2014MNRAS.445..581H,Agertz13,2015ApJ...804...18A} inject HII region momentum via a similar prescription to ours, but in the analytic form of a `direct radiation pressure' during the radiation-dominated phase of HII region expansion. As discussed in KM09 and in our Section~\ref{Sec::Theory}, radiation pressure dominates the expansion of only the largest HII regions, while those with ionising luminosities in the range of $1 < S_{49} < 100$ (as for the FoF groups in our high-resolution simulation) suffer a factor of ten or more reduction in the momentum injected, if the gas-pressure term in Equation~(\ref{Eqn::momentumeqn_final}) is ignored. Despite this, the above works find that their radiation pressure prescriptions are necessary to achieve a realistic interstellar medium. This can be attributed to their use of an $f_{\rm trap}$ factor far exceeding that found in observations~\citep[e.g.][]{2021ApJ...908...68O}. In~\cite{2013MNRAS.434.3142A,Agertz13,2015ApJ...804...18A} a value of $f_{\rm trap} \sim 25$ is used, and in~\cite{2011MNRAS.417..950H,2014MNRAS.445..581H} this value is further increased within the range $f_{\rm trap} \sim 10$-$100$. By contrast, later works of~\cite{Hopkins18,2019MNRAS.489.4233M} reduce the value of $f_{\rm trap}$ back to order one, and the authors find that in this case, the direct radiation pressure has a negligible effect~\citep[see Figure 36 of][]{Hopkins18}. In summary, the above works agree with our results in the sense that a pre-supernova momentum injection of $\sim 10$-$100 \times L/c$ has a substantial influence on the intermediate- and large-scale properties of the interstellar medium. This momentum injection is needed to achieve an interstellar medium consistent with observations. However, according to the calculations presented in KM09 and in our Section~\ref{Sec::Theory}, the vast majority of this momentum comes from the gas pressure inside the HII region, and not from the radiation pressure.

Finally, we note that an identical feedback prescription (but using $f_{\rm trap} = 2$ rather than $f_{\rm trap} = 8$) was adopted in~\cite{2020MNRAS.498..385J} and in~\cite{Jeffreson21a} to investigate molecular cloud properties in three isolated disc galaxies with external, analytic galactic potentials. The molecular cloud population at the native resolution in these studies was on average less massive (maximum mass of $\sim 10^7~{\rm M}_\odot$ vs. $\sim 10^8~{\rm M}_\odot$ here) and had a shorter median cloud lifetime ($\sim 20$~Myr vs. $\sim 35$~Myr here). This can be attributed to the fact that the mid-plane turbulent pressure in the Agora disc used here is approximately eight times that of the discs introduced in~\cite{2020MNRAS.498..385J}, i.e.~the mid-plane gas surface density and velocity dispersion are both doubled. This may be due to the use of a live dark matter and stellar potential, which allows for a greater degree of baryon clustering. The star formation efficiency per free-fall time of $\epsilon_{\rm ff} = 10$~per~cent used in this work is also ten times the value of $1$~per~cent used in~\cite{2020MNRAS.498..385J}, because we have found that the lower star formation efficiency results in unphysically-bursty star formation, and an unphysically-high turbulent velocity dispersion of the cold gas on kpc-scales.

\section{Conclusions} \label{Sec::conclusion}
In this work, we have developed a novel model for the momentum imparted by marginally-resolved HII regions in simulations with mass resolutions between $10^3$ and $10^5~{\rm M}_\odot$ per gas cell. The model can be applied in a spherical or a beamed configuration, where the latter corresponds to the directed momentum injected from blister-type HII regions on the edges of molecular clouds. We have compared simulations with only supernova feedback to simulations with supernova and thermal HII region feedback, spherical HII region momentum, beamed HII region momentum, and a combination of beamed momentum and thermal injection, across the mass resolution range $10^3$-$10^5~{\rm M}_\odot$. In general, we find that:
\begin{enumerate}
	\item Thermal feedback from marginally-resolved HII regions has no influence on the interstellar medium, at any scale or resolution.
	\item The geometry of momentum injection (spherical or beamed) from HII regions similarly has very little effect.
\end{enumerate}
When HII region momentum is introduced at mass resolutions between $10^3$ and $10^4~{\rm M}_\odot$, the large-scale interstellar medium responds in the following ways:
\begin{enumerate}
	\item The mass-loading and magnitude of galactic outflows are reduced by an order of magnitude.
	\item The gas-disc scale-height is reduced by $0.5$~dex for galactocentric radii $>5$~kpc.
	\item The velocity dispersion of the cold gas is supressed by $\sim 5~{\rm kms}^{-1}$, and the gravitational stability of the gas disc is correspondingly decreased  by a factor of around two in the~\cite{Toomre64} $Q$ parameter.\footnote{We recall that the factors of decrease in the velocity dispersion and Toomre $Q$ parameter do not match exactly because of the difference in gas density at the simulation time of comparison.}
  \item The mass fraction of cold gas is increased by $\sim 50$~per~cent and the mass fraction of cold molecular gas is approximately doubled.
\end{enumerate}
The above results are consistent with the idea that HII region feedback (and pre-supernova feedback in general) reduces the clustering of supernovae and therefore their efficacy in depositing large quantities of momentum into the interstellar medium, as studied by~\cite{2020arXiv200911309S}. The results do not persist down to resolutions of $10^5~{\rm M}_\odot$, although the HII regions are still marginally-resolved. This non-convergence is likely due to a decrease in the effectiveness of supernova clustering.

For molecular clouds specifically, we find the following results:
\begin{enumerate}
	\item The lifetimes of the least massive molecular clouds ($M/{\rm M}_\odot \la 5.6 \times 10^4$) are reduced from $\sim 18$~Myr to $<10$~Myr. That is, HII region momentum is able to disperse low-mass clouds that do not contain supernovae from massive stellar clusters.
	\item The lifetimes of intermediate-mass ($5.6 \times 10^4 \la M/{\rm M}_\odot \la 5 \times 10^5$) clouds are increased by $\sim 7$~Myr. That is, HII region momentum decreases the efficiency of supernovae in dispersing intermediate-mass clouds.
	\item The molecular cloud star formation rate surface density is suppressed by a factor of three.
	\item The molecular cloud velocity dispersion is reduced by $\sim 0.5~{\rm kms}^{-1}.$
\end{enumerate}

In summary, we find that the large- and intermediate-scale properties of the simulated interstellar medium, as well as the properties of its molecular clouds, are significantly altered by the introduction of momentum from HII regions at numerical mass resolutions from $10^3~{\rm M}_\odot$ to $10^5~{\rm M}_\odot$ per gas cell. More than 90~per~cent of the injected momentum is due to the thermal expansion of the HII regions, rather than radiation pressure. The injection of thermal energy without momentum has no discernible effect on the simulated galaxies.

\section*{Acknowledgements}
We thank an anonymous referee for a constructive report, which improved the paper. We thank Volker Springel for providing us access to Arepo. SMRJ is supported by Harvard University through the ITC. We gratefully acknowledge funding from the Deutsche Forschungsgemeinschaft (DFG, German Research Foundation) through an Emmy Noether Research Group (SMRJ, MC, JMDK; grant number KR4801/1-1) and the DFG Sachbeihilfe (MC, JMDK; grant number KR4801/2-1), as well as from the European Research Council (ERC) under the European Union's Horizon 2020 research and innovation programme via the ERC Starting Grant MUSTANG (SMRJ, BWK, JMDK; grant agreement number 714907). SMRJ, MRK, YF, MC and JMDK acknowledge support from a UA-DAAD grant. BWK acknowledges funding in the form of a Postdoctoral Research Fellowship from the Alexander von Humboldt Stiftung. MRK acknowledges support from the Australian Research Council through Future Fellowship FT80100375 and Discovery Projects award DP190101258. The work of LA was partly supported by the Simons Foundation under grant no.~510940. The work was undertaken with the assistance of resources and services from the National Computational Infrastructure (NCI; award jh2), which is supported by the Australian Government. We are grateful to Jeong-Gyu Kim, Eve Ostriker and Vadim Semenov for helpful discussions.

%%%%%%%%%%%%%%%%%%%%%%%%%%%%%%%%%%%%%%%%%%%%%%%%%%
\section*{Data Availability Statement}
The data underlying this article are available in the article and in its online supplementary material.



\bibliographystyle{mnras}
\bibliography{bibliography} % if your bibtex file is called example.bib

%%%%%%%%%%%%%%%%%%%%%%%%%%%%%%%%%%%%%%%%%%%%%%%%%%

%%%%%%%%%%%%%%%%% APPENDICES %%%%%%%%%%%%%%%%%%%%%

\appendix
\section{Convergence test for supernovae} \label{App::res-tests}
Here we examine the convergence of the energy and momentum injected by a single supernova explosion in our simulations, as a function of the median gas cell mass. The prescription for momentum injection is outlined in Section~\ref{Sec::SNe-only}. As discussed in Section~\ref{Sec::grouping} for the case of the HII region mass and luminosity, deviations from convergence in the final simulations may still arise due to resolution-dependent variations in the clustering of star particles and their supernovae.

\begin{figure}
  \label{Fig::SNe-convergence}
    \includegraphics[width=\linewidth]{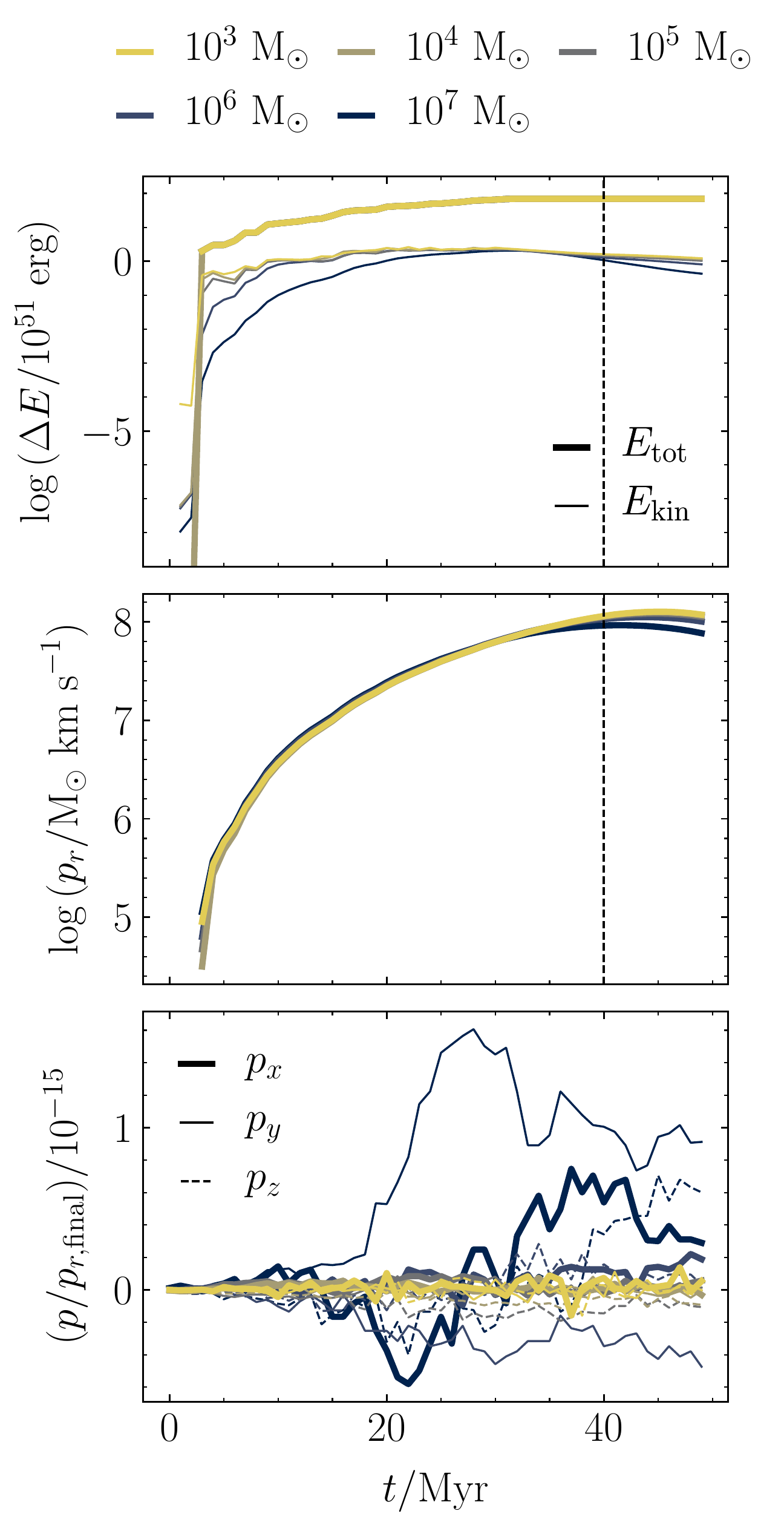}
    \caption{Energy (top panel) and radial momentum (central panel) of the gas cells in a box of size $(950~{\rm pc})^3$ and density $100~{\rm cm}^3$ containing a single star particle of mass $1 \times 10^4~{\rm M}_\odot$, as a function of time. The star particle injects SN feedback according to the prescription outlined in Section~\ref{Sec::SNe-only}. The resolution varies from a lowest resolution of $10^7~{\rm M}_\odot$ per gas cell (dark blue) up to a highest resolution of $10^3~{\rm M}_\odot$ per gas cell (yellow). The linear momentum is conserved at all resolutions (lower panel). The vertical dashed black line represents the time at which feedback is switched off for each star particle in our simulations (the SLUG object is deleted).}
\end{figure}

In Figure~\ref{Fig::SNe-convergence}, we show the radial momentum and energy from supernovae injected by a single star particle over a time period of $40$~Myr. The feedback is injected into a box of size $(950~{\rm pc})^3$ and density $100~{\rm cm}^{-3}$ at mass resolutions varying between $10^7~{\rm M}_\odot$ per gas cell ($8^3$ cells) and $10^3~{\rm M}_\odot$ per gas cell ($128^3$ cells). The top panel shows that the kinetic (thin lines) and total (bold lines) energies are converging with resolution, while the central panel shows the same result for the radial momentum of the gas cells away from the central star particle. The lower panel demonstrates that linear momentum is conserved to machine precision.

%========================= SNE CONVERGENCE FIG
%=========================

\section{Thermal and chemical post-processing} \label{App::postproc}
In order to calculate the mass fractions of molecular gas used to produce Figures \ref{Fig::vss}, \ref{Fig::KS-rln}, \ref{Fig::cloud-lifetimes}, \ref{Fig::cloud-veldisp-surfdens}, and \ref{Fig::cloud-mass-SFR}, we post-process the physical fields in our simulations using the astrochemistry and radiative transfer model {\sc Despotic}~\citep{Krumholz14}. Following the method used in~\cite{Fujimoto19,2020MNRAS.498..385J}, we treat each Voronoi gas cell as a one-zone spherical `cloud', with characteristic values of the hydrogen number density, column density and virial parameter. The line emission from each cloud model is calculated within {\sc Despotic} via the escape probability formalism, which is self-consistently coupled to the carbon, oxygen and hydrogen chemistry of the gas (followed via the chemical network of~\citealt{Gong17}), and to the thermal evolution of the gas (including heating by cosmic rays, the grain photo-electric effect, line cooling due to ${\rm C}^+$, ${\rm C}$, ${\rm O}$, and ${\rm CO}$, and the thermal exchange between dust and gas). The strength of the interstellar radiation field and the cosmic ionisation rate are set equal to the values used during the computation of live chemistry. The computation yields the $^{12}{\rm CO}$ line luminosity $L_{\rm CO}$ for the $J=1 \rightarrow 0$ transition~\citep{1977JPhB...10..879L,1987JPhB...20.2553J,1991JPhB...24.2487S,1998MNRAS.293L..83B,2002MNRAS.337.1027W,2005A&A...432..369S,2005ApJ...620..537B,2010ApJ...718.1062Y,2013JChPh.138t4314L}.

It would be prohibitively computationally-expensive to perform the described convergence calculation for every gas cell in the simulation, so instead we interpolate over a table of models at logarithmically-spaced intervals in the volume density, column density and virial parameter. The column density is computed by applying the local Jeans length approximation of~\cite{Safranek-Shrader+17} to the volume density of each gas cell, and the virial parameter is defined according to~\cite{MacLaren88,BertoldiMcKee1992}. We then use the two-dimensional projection of $L_{\rm CO}$ to compute the CO-bright molecular hydrogen surface density, as
\begin{equation} \label{Eqn::mol-col-dens}
\begin{split}
\Sigma_{\rm H_2}[{\rm M}_\odot {\rm pc}^{-2}] = &\frac{2.3 \times 10^{-29}[{\rm M}_\odot ({\rm erg}~{\rm s}^{-1})^{-1}]}{m_{\rm H}[{\rm M})_\odot]} \\
&\times \int^\infty_{-\infty} \dd z^\prime \rho_{\rm g}(z^\prime) L_{\rm CO}[{\rm erg}~{\rm s}^{-1}],
\end{split}
\end{equation}
where $\rho_{\rm g}$ is the gas volume density as a function of the perpendicular distance from the galactic mid-plane, $\Sigma_{\rm g}$ is the gas surface density, and $m_{\rm H}$ is the proton mass. The factor $2.3 \times 10^{-29} ({\rm erg~s}^{-1})^{-1}$ combines the mass-to-luminosity conversion factor of $\alpha_{\rm CO} = 4.3~{\rm M}_\odot~({\rm K}~{\rm kms}^{-1}{\rm pc}^{-2})^{-1}$ from~\cite{Bolatto13} and the line-luminosity conversion factor of $5.31 \times 10^{-30}({\rm K}~{\rm kms}^{-1}{\rm pc}^2)/({\rm erg}~{\rm s}^{-1})$ from~\cite{SolomonVandenBout05} for the CO $J=1\rightarrow 0$ transition at redshift $z = 0$. The integrals are performed using the in-built ray-tracing modules in {\sc Arepo}.

Once we have calculated maps of the molecular hydrogen column density at our native resolution, according to Equation~\ref{Eqn::mol-col-dens}, we use the {\sc Astrodendro} package for Python to pick out molecular clouds as the set of closed contours at $\log{(\Sigma_{\rm H_2}/{\rm M}~{\rm pc}^{-2})} = -3.5$. The choice of such a lenient threshold corresponds to a natural break in the molecular hydrogen column density distribution produced by the chemical post-processing in {\sc Despotic}. Above the threshold, we find gas cells that contain at least some UV-shielded, CO-dominated gas, and below the threshold we find cells for which the CO and molecular hydrogen are unshielded and uniformly-mixed, with very low overall abundances. We find this a preferable alternative to taking an arbitrary, higher-density cut-off, as discussed in~\cite{2020MNRAS.498..385J,Jeffreson21a}. However, the choice does not matter hugely, as most of the CO resides in gas cells at much higher density. We find that increasing the threshold to $10~{\rm M}_\odot~{\rm pc}^{-2}$ changes the total area of the molecular cloud sample by less than $5$~per~cent. By applying the {\sc Astrodendro} pixel masks to the field of Voronoi gas cell centroids with temperatures $T < 10^4$~K, we obtain the gas distribution within each identified cloud.

%%%%%%%%%%%%%%%%%%%%%%%%%%%%%%%%%%%%%%%%%%%%%%%%%%

% Don't change these lines
\bsp	% typesetting comment
\label{lastpage}
\end{document}